\documentclass[letterpaper]{JHEP}

\usepackage{float}
\usepackage{feynmp}
\usepackage[hoffset=0in, voffset=0in]{geometry}
\usepackage{amsmath}
\usepackage{amssymb}
\newcommand{\slashed}[1]{#1\!\!\!\!/ \,\,}
\usepackage[bf,small]{caption}
\usepackage{bbold}
\newcommand{\eqn}[1]{(\ref{#1})}

\title{Spinors on manifolds with boundary: APS index theorems with torsion}
\author{Kasper Peeters\\
DAMTP, Cambridge University\\
Silver Street\\
Cambridge CB3 9EW\\
United Kingdom\\
\email{k.peeters@damtp.cam.ac.uk}}
\author{Andrew Waldron\\
NIKHEF\\
P.O. Box 41882\\
1009 DB Amsterdam\\
The Netherlands\\
\email{waldron@nikhef.nl}}
\keywords{index theorems, supersymmetry, Taub--NUT, Killing--Yano, torsion, path integrals}

\abstract{Index theorems for the Dirac operator allow one to study
spinors on manifolds with boundary and torsion. We analyse the
modifications of the boundary Chern--Simons correction and APS $\eta$
invariant in the presence of torsion. The bulk contribution must also
be modified and is computed using a supersymmetric quantum mechanics
representation. Here we find agreement with existing results which
employed heat kernel and Pauli--Villars techniques.  Nonetheless, this
computation also provides a stringent check of the Feynman rules of de
Boer {\it et al.}~for the computation of quantum mechanical path
integrals. Our results can be verified via a duality relation between
manifolds admitting a Killing--Yano tensor and manifolds with torsion.
As an explicit example, we compute the indices of Taub--NUT and its
dual constructed using this method and find agreement for any finite
radius to the boundary. We also suggest a resolution to the
problematic appearance of the Nieh--Yan invariant multiplied by the
regulator $(\rm mass)^2$ in computations of the chiral gravitational
anomaly coupled to torsion.}

\preprint{DAMTP-1998-169, NIKHEF 98-034, \hepth{9901016}}

\begin{document}
\section{Introduction}

Index theorems for manifolds with boundary and torsion involve a
detailed understanding of spinors on such manifolds. In the torsion
free case this problem was understood long ago in the mathematics
literature and is solved by the Atiyah--Patodi--Singer index
theorem~\cite{atiy1}.  Essentially, they found that the index for
manifolds with boundary is the sum of the usual bulk result for the
index (proportional to the integral over the first Pontrjagin class)
plus a correction from the so-called $\eta$ invariant of the boundary
Dirac operator.  Upon the addition of torsion, further questions arise
such as how does torsion affect the computation of the index in the
bulk?  What are the appropriate boundary conditions for spinors on
such manifolds and how does one define the adjoint of operators acting
on these spinors?  How does torsion modify the computation of the
$\eta$ invariant? Are there corrections to the index over and above
that found by Gilkey~\cite{gilk1} if the metric is not of product form
near the boundary? {}From a physical viewpoint the resolution of these
questions is of considerable interest, not only because of the
relation between the index of the Dirac operator and anomalies, but
also due to the considerable importance attached to the physics of
supergravity theories (involving, of course, both spinors and torsion)
on manifolds with boundary, {\it e.g.}~anti de Sitter spaces~\cite{mald2}.

Manifolds in the Einstein--Cartan theory which include torsion are not
nearly as well studied as those without, and solutions of the
equations of motion tend to be of considerable complexity, making
explicit checks of the general formalism we are about to present
cumbersome. Fortunately, we can relate the index of a manifold ${\cal
M}$ with torsion to a torsion free manifold $\widetilde{\cal M}$ when
their Dirac operators, $\slashed{D}(e,A)$ and $\slashed{D}(\tilde e,\widetilde V)$
respectively, satisfy
\footnote{The existence of a relation of this type was first
suggested by Jan-Willem van Holten
based on the vanishing classical Poisson brackets between the  supercharges of
manifolds with Killing--Yano supersymmetries~\cite{holt5}.}
\begin{equation}
\label{e:anti}
\{\,\slashed{D}(e,A),\slashed{D}(\tilde e,\widetilde V)\,\}\,=\,0
\end{equation}
(note that $\slashed{D}(\tilde e,\widetilde V)$ denotes the Dirac operator
with inverse vierbein $\tilde e_r{}^\mu$ coupled minimally to an
abelian vector field $\widetilde V^\mu$, whose necessity will be explained
later, and $\slashed{D}(e,A)$ refers to inverse vierbein $e_r{}^\mu$
and coupling to totally antisymmetric torsion which may be
re-expressed via Hodge duality as minimal coupling to an axial vector
field $A^\mu$).  This equation will be made mathematically more
precise later, it suffices to say here that if we are able to find a
pair of manifolds whose Dirac operators are related in this way, we
will be able to test our results for index theorems on manifolds with
boundary in a concrete example. In what follows we will refer to the
manifold with torsion $\cal M$ as the ``dual'' of the torsion-less
manifold $\widetilde{\cal M}$.

In fact, as was pointed out in \cite{riet2}, for the case where one
takes the manifold $\widetilde{\cal M}$ as the background for
classical spinning particle models with extended supersymmetry, one
may view $\slashed{D}(\tilde e,\widetilde V)=\gamma^r \widetilde
e_r{}^\mu\partial_\mu+\cdots$ as a supersymmetry generator and then
search for additional supersymmetry generators
$\slashed{D}(e,A)=\gamma^r e_r{}^\mu\partial_\mu+ \cdots$. The tensor
$e_r{}^\mu$ is known as a Killing--Yano tensor and may be viewed as
the inverse vierbein of the dual manifold ${\cal M}$. We also
stress that in the quantum case, when one considers the Dirac
operators $\slashed{D}(e,A)$ and $\slashed{D}(\tilde e,\widetilde V)$
depending on Dirac matrices satisfying a Clifford algebra,
equation~\eqn{e:anti} is more subtle than in the classical case where
one has supersymmetry charges depending on world line Majorana spinors
obeying canonical Poisson brackets. The quantisation of Killing--Yano
symmetries is, in fact, an important aspect of our work.

At the end of this paper we take as an example
$\widetilde {\cal M}$ to be Taub--NUT, 
\begin{align}
\label{e:taub-nut_metric}
d\tilde s^2 &=\frac{r+2m}{r}\,\Big[ {\rm d}r^2 + r^2\, {\rm d}\theta^2 +
    r^2\sin^2\theta\, {\rm d}\phi^2\Big]+ \frac{4r m^2}{r+2m}\,\Big[
    {\rm d}\psi + \cos\theta\, {\rm d}\phi\Big]^2\\
\intertext{a manifold of considerable physical interest and ${\cal M}$ is then 
the dual Taub--NUT manifold with metric}
\label{e:dual_metric}
{\rm d}s^2 &= 
    \frac{r+2m}{r}\Big[ {\rm d}r^2
   +\frac{m^2r^2}{(r+m)^2} \left( {\rm d}\theta^2 
        + \sin^2\theta\,{\rm d}\phi^2 \right) \Big] 
   + \frac{4rm^2}{r+2m} \Big[ {\rm d}\psi + \cos\theta\,{\rm d}\phi\Big]^2\, ,
\end{align}
along with totally antisymmetric torsion (our normalisations can be
found in the appendix)
\begin{equation}
T=-\frac{4r^2 m^2 \sin \theta}{(r+m)^2}\, {\rm d}\theta\wedge {\rm
d}\phi\wedge {\rm d}\psi\, .
\end{equation}
We should mention that the dual Taub--NUT manifold possesses neither
self-dual Riemann curvature tensors nor satisfies the Einstein--Cartan
equations but this is unimportant for our purposes since it does
satisfy the central equation~\eqn{e:anti}.

Our chief line of argument to compute the index of a manifold with
torsion and boundary follows the Atiyah--Patodi--Singer index
theorem~\cite{atiy1}.  We split the computation into a contribution
from the bulk and a boundary contribution. The bulk contribution is
computed by representing the index as the regulated trace of
$\gamma^5$
\begin{equation}
\label{e:elvis}
{\rm index} \,  \slashed{D}\,=\,{\rm Tr}\left[ \gamma^5\,\exp
\Big(\frac{\textstyle\beta}{\textstyle2}\slashed{D}^2\Big)\right]
\end{equation}
which in turn may be represented as a supersymmetric quantum
mechanical path integral~\cite{alva15}.  In this approach, however,
one essentially inserts infinitely many sets of plane wave resolutions
of unity, for which one may clearly integrate by parts. Therefore this
calculation is insensitive to the precise boundary conditions obeyed
by the states appearing in the trace in~\eqn{e:elvis} and yields only
the bulk contribution to the index. In fact, analogous computations
have also been carried out in a heat kernel~\cite{obuk2} as well as
Pauli--Villars~\cite{doba1} framework and yield results identical to
ours. Nonetheless, our independent computation is interesting in its
own right since it provides a stringent test of the rigorous quantum
mechanical path integral formalism for sigma models in curved space
developed by de Boer {\it et al.}~\cite{boer4,boer5} (and in
particular of the Christ and Lee $\hbar^2$ counter terms~\cite{chri3}
that must be added to the action in order that the path integral
provides a faithful representation of the trace~\eqn{e:elvis}).  The
details of this computation are described in section~\ref{s:bulk}.

Having calculated the bulk contribution one must compute the boundary
contribution which has two parts. The first is the $\eta$ invariant
for the boundary manifold. The $\eta$ invariant is a spectral
invariant of the boundary Dirac operator and may be computed from the
eigenvalues of this operator. In the torsion free case this problem
was solved for various asymptotic geometries by Hitchin (and applied
to the Taub--NUT manifold in~\cite{roem1,eguc4,pope1}). Therefore, 
we generalise the computation of Hitchin to include the
dual Taub--NUT example. To this end we construct and solve the boundary Dirac
operator in the presence of torsion which allows us to compute the
$\eta$ invariant for this manifold.

The second boundary contribution, due to Gilkey~\cite{gilk1}, is
necessary in the case where the manifold is not of product form at the
boundary. The main idea is that the APS analysis is valid for
manifolds which approach a cylinder whose cross section is the
boundary manifold. If this is not the case, one must make an extra
boundary correction proportional to the integral over the boundary of
the difference between the Chern--Simons form computed using the
product metric and that computed using the metric itself. This
computation can be easily generalised to the torsion-full case where
we find a boundary correction proportional to a generalised
Chern--Simons term.  Note, in particular, that all our results hold
for manifolds with boundaries at any finite radius, rather than just
in some non-compact limit. The derivations of the boundary corrections
are to be found in section~\ref{s:boundary}.

Section~\ref{s:dual_operators} analyses in detail the precise relation
between index theorems of manifolds possessing mutually anticommuting
Dirac operators. This relation requires a careful understanding of
boundary conditions for spinors and therefore the space of states in
the trace~\eqn{e:elvis}.  Subsection~\ref{s:ky_duality} outlines the
main ingredients of the Killing--Yano technique to generate manifolds
with mutually anticommuting Dirac operators and the explicit geometric
data for our example manifolds, Taub--NUT and dual Taub--NUT may be
found in~\ref{ss:taub-nut}.  Also included in
subsection~\ref{s:ky_duality} is the generalisation of the
Killing--Yano construction to the quantum case.

Urgent readers not wishing to wade through all the details may find
the final expression for the index with torsion on manifolds with
boundary in equation~\eqn{e:johnny}. A summary of the results we
obtained for the example manifolds is presented in
subsection~\ref{ss:example_indexsummary}. Finally, more speculative
remarks along with possible applications of our results are reserved
for the conclusion where, in particular, we discuss the controversial
$1/\beta$ regulator $(\mbox{mass})^2$ Nieh--Yan contribution to the
index in light of our work.  In passing, we note that many of the
issues we encounter have been dealt with in some depth in the
mathematics and physics literature so that we have attempted to
present these more formal aspects in an accessible format.

\section{Bulk contribution to the index with torsion}
\label{s:bulk}
\subsection{The Dirac index as a quantum mechanical path integral}

This section is devoted to the computation of the bulk computation to
the index employing the supersymmetric path integral approach of
Alvarez-Gaum\'e and Witten~\cite{alva15}.  The bulk contribution to
the index of the Dirac operator with torsion has already been computed
in~\cite{obuk2} via a heat kernel expansion and in~\cite{doba1} using
Pauli--Villars regularisation, needless to say, our results exactly reproduce
those of these authors. Besides the high degree of certainty one now
has in the result (see equation~\eqn{e:bulkresult}), our motivations
for performing this calculation were fourfold:
\begin{enumerate}
\item
Our computation
is carried out in the framework of rigorously defined quantum mechanical
curved space path integrals introduced by de Boer {\it et al.}~\cite{boer4,boer5}
and therefore provides a stringent check of their work. In particular,
we note that the result depends crucially on the Christ and Lee 
terms~\cite{chri3} (see the second line of~\eqn{e:action}) that
one must add to the action appearing in the path integral at order $\hbar^2=\beta^2$
in order to employ propagators derived via a midpoint rule. 
\item
There has been some controversy in the literature over the appearance
of a term in the (bulk) index with torsion that diverges as one takes the
limit in which the regulator is
removed~\cite{obuk1,soo1,miel1,chan4,chan5,chan6}. This term is proportional
to the Nieh--Yan tensor (see the first term on the right hand side
of~\eqn{e:bulkresult}) and arises at tree level in our path integral
expansion.  In the path integral approach, one clearly sees the
importance of including this term in order to obtain the correct
results for the bulk index ({\it i.e.}~the regulator independent
contribution) since it appears as a vertex as well as in disconnected
graphs.
\item 
The path integral derivation is based on inserting infinitely many plane wave
resolutions of unity. Therefore one clearly sees that it is only sensitive to the bulk
contribution, since differentiations by parts then ignore any boundary terms.
\item
The quantum mechanical 
model itself is of intrinsic interest since it describes a spinning particle
moving through a background with torsion. Obviously, the construction of the
precise path integral quantisation of this model is an important step
in understanding its quantum dynamics.
\end{enumerate}
The index of the Dirac operator $\slashed{D}$ is given by the
trace over the spectrum
\begin{equation}
\label{e:trace}
{\rm index}\, \slashed{D} = \text{Tr}\left[ \gamma^5\,\exp
\Big(\frac{\textstyle\beta}{\textstyle2}\slashed{D}^2\Big)\right]\, .
\end{equation}
When there is a gap between the zero-modes and the remainder of the
spectrum~\cite{akho1} this result should, of course, be $\beta$-independent. 
However, for manifolds with boundary, one must carefully
take into account the states traced over. If one is interested in
the bulk contribution only, the trace can be evaluated by inserting $N$
sets of plane waves (where ${\rm d}t\sim 1/N\rightarrow 0$) and
the result may then be represented as a path integral. The expression one
then finds is an expansion in powers of $\beta$
\begin{equation}
\beta^{-1}\Delta_{-1}(\slashed{D})+\Delta_0(\slashed{D})+O(\beta)
\,\,\equiv\,\,\Delta(\slashed{D}) 
\, .
\end{equation}
The one-loop beta independent contribution $\Delta_0(\slashed{D})$
yields the bulk result for the index and as will be explained in
section~\ref{s:boundary}, the complete result for the index is
obtained by including the APS and Chern--Simons boundary correction
terms.  Here we will compute the terms $\Delta_{-1}(\slashed{D})$ and
$\Delta_0(\slashed{D})$. We comment on the interpretation of
the $1/\beta$ contribution $\Delta_{-1}(\slashed{D})$ in the
conclusion.

Before introducing the details of our path integral computation, let us briefly
summarise the main ingredients of the approach. To begin with, one 
treats the exponential in the trace~\eqn{e:trace} 
as the imaginary time evolution operator of a quantum 
mechanical system. The {\em operator\/}
in the exponent corresponds to the Hamiltonian.  
Importantly, its operator ordering is now fixed so that the usual step
where one begins with some classical Hamiltonian generating dynamics via 
Poisson brackets and then elevates this expression to an operator, fixing the
ordering by symmetry principles (or ultimately experiment!) is not needed.
The operator $H=-\beta^2\slashed{D}^2/2$ {\em is\/} the {\em quantum}\/ Hamiltonian.

The next step is to derive the path integral by inserting $N$
complete sets
of states. The achievement of de Boer {\it et al.}~was to derive a path integral
which faithfully represents the operator expression~\eqn{e:trace}. 
The point being that once one takes the continuum limit $N\rightarrow\infty$,
the discretised expressions derived at finite $N$ become distributions whose
products (which appear in Feynman diagrams) are ambiguous. 
However, given the rigorous discretised expression
for the path integral it is always possible to resolve these ambiguities
in a unique and finite fashion. Their derivation was performed by {\em rewriting\/}
the quantum Hamiltonian in a Weyl ordered form which allows infinitesimal
transition elements at intermediate steps in the path to be evaluated by a midpoint 
rule which leads to a certain set of vertices, propagators and rules for
products of distributions. Of course another ordering principle would have yielded
different propagators and vertices, the key point is to derive a path integral
which precisely 
represents\footnote{It is interesting to compare this situation 
to what one finds in
quantum field theory. In quantum field theory, one must regulate
and renormalise to obtain a path integral yielding finite, well defined
results for the products of distributions appearing at loop level in Feynman
diagrams. 
Furthermore, one makes some choice of renormalisation point
in order to make contact with physical quantities. 
Of course, in quantum 
mechanics, there is no anomaly in scale invariance and the path integral
$Z[y^\mu,\Psi^r]$ (see~\eqn{e:zeZ}) should provide an unambiguous and finite
representation of the transition amplitude 
$\langle y,\Psi|\,\gamma^5\,\exp \frac{\beta}{2}\slashed{D}^2\,|y,\Psi\rangle$.
Although, the quantum mechanical
Hamiltonian $H=-\beta^2 \slashed{D}^2/2$ has a definite
operator ordering, in passing to a path integral representation, the 
vertices of the path integral are 
ambiguous up to reorderings of the Hamiltonian.
However, once a particular ordering 
(there is nothing sacrosanct about Weyl ordering, it is merely convenient)
is fixed one must decide (derive)
which rule is used to define products of distributions. In analogy with
quantum field theory theory, the discretisation procedure and rule for the 
products of delta and Heaviside functions above 
may be thought of as regulating
and renormalising the theory with renormalisation point judiciously chosen 
to reproduce the required transition amplitude with the 
given Hamiltonian.}~\eqn{e:trace}. 

We now turn to  the details our calculation 
and begin by  thinking  of the exponential in~\eqn{e:trace}
\begin{equation}
\exp[-i(-\beta\,^2 \slashed{D}^2/2)(-i)/\beta]
\end{equation} 
as a quantum mechanical 
evolution operator $\exp[-iHt/\hbar]$ with Hamiltonian 
$-\beta\,^2 \slashed{D}^2/2$,
unit Euclidean time interval and Planck's constant $\hbar=\beta$.
The Dirac matrices are then identified with fermionic coordinates
and the derivative operator $\partial_\mu$ with the canonical momentum
of the bosonic coordinate $x^\mu$,
\begin{equation}
\label{e:identify}
\begin{aligned}
\gamma^r=\sqrt{2}\,\beta^{-1/2}\,\psi^r\,, &\quad\quad 
\partial_\mu=\beta^{-1}g^{\frac{1}{4}}\, ip_\mu\, g^{-\frac{1}{4}}\,,\\[.5em]
\{\psi^r,\psi^s\}=\beta\,\delta^{rs}\,, &\quad\quad
[x^\mu,p_\nu]=i\beta\,\delta^{\mu}{}_\nu\, .
\end{aligned}
\end{equation}
The similarity transformation for the canonical momentum $p_\mu$
(with $g=\det g_{\mu\nu}$) implies
the inner product $\langle x | y \rangle=g^{-1/2}\delta^4(x-y)$.
Observe also that the rescalings by appropriate factors of $\beta$
in~\eqn{e:identify} ensure that $\beta$ plays the r\^ole of Planck's
constant in the canonical commutation relations and that
the Hamiltonian $-\beta\,^2 \slashed{D}^2/2$ begins with a ``classical''
$\beta$ independent contribution, plus possibly terms linear and
quadratic in $\beta$ (depending on the precise ordering in which one
writes the operators $x^\mu$, $p_\mu$ and $\psi^r$).

We may now represent the index~\eqn{e:trace} by a path integral with
periodic boundary conditions, the presence of $\gamma^5$ in the trace
yields periodic boundary conditions for the fermions also.
\begin{equation}
\label{e:pathrep}
\Delta(\slashed{D})=\int_{\cal M} \! {\rm d}^4y\, \frac{\sqrt{g(y)}}{(2\pi i)^2}
\, {\rm d}^4\Psi\, Z[y^\mu,\Psi^r]
\end{equation}
where the path integral $Z$ is a function of constant real background
fields $y^\mu$ and $\Psi^r$. Schematically
\begin{equation}
\label{e:zeZ}
\begin{aligned}
Z[y^\mu,\Psi^r]=\int&[{\rm d}q^\mu\, {\rm d}\eta^r\, {\rm d}a^\mu
\,{\rm d}b^\mu \,{\rm d}c^\mu]\\
& \exp\Big\{-\frac{1}{\beta}\int_0^1\!{\rm d}t\, {\cal L}
(x^\mu(t)=y^\mu+q^\mu(t),\psi^r(t)=\Psi^r+\eta^r(t))\Big\}\, .
\end{aligned}
\end{equation}
As discussed above, one usually expects path integrals to be plagued
with ambiguities associated with the normalisation and precise
definition of the measure along with products of distributions in
loops.  Fortunately however, finite interval path integrals in curved
space have been studied in detail by de Boer {\it et
al.}~\cite{boer4,boer5}. In particular they have found the precise
definition of $Z$ above, {\it i.e.}~the exact vertices and Feynman
rules, such that the path integral representation~\eqn{e:pathrep} is
identically equal to the index $\Delta(\slashed{D})$. A two loop
verification of their work in the torsion-less case may be found
in~\cite{wald1} along with the conventions employed in this paper.  We
now spell out the key details of this approach.

Firstly, although manifestly the 
quantum mechanical trace~\eqn{e:trace} is finite, in the path integral,
closed bose loops $\langle \dot q^\mu(t)\dot q^\nu(t)\rangle$
yield delta function divergences $\delta(0)$ (first observed by
Lee and Yang~\cite{lee2}). However, the decomposition of unity
$1=\int\!{\rm d}^4x\, g^{1/2}|x\rangle\, \langle x|$ yields a measure
factor $g^{1/2}$ at each point of the path which one may exponentiate
via bosonic and fermionic ghosts $a^\mu(t)$ and $\{b^\mu(t),c^\mu(t)\}$,
respectively~\cite{bast2,bast3} yielding the ghost action
\begin{equation}
\int_0^1\!{\rm d}t\,
{\cal L}_{\rm ghost}=\int_0^1\! {\rm d}t\,
\frac{1}{2}g(x(t))_{\mu\nu}\big(a(t)^\mu a(t)^\nu+2\,b(t)^\mu c(t)^\nu\big)\, ,
\end{equation}
whose net effect is to precisely cancel the divergence $\delta(0)$
whenever it appears (for example graphs including a ghost loop are needed
to cancel the divergence in the third graph of table~\ref{t:twoloop}).

Secondly, ambiguous products of distributions involving Heaviside and
delta functions $\delta(t-s)\theta(t-s)$ also appear. The result
of~\cite{boer4,boer5} is that adopting vertices corresponding to a
{\em Weyl ordered Hamiltonian}, one finds propagators such that
$\theta(0)=1/2$ (Strictly speaking, one returns to the discretised
derivation of the path integral where the delta function is a
Kronecker delta $\delta_{ij}$ and the propagators depend on the
discrete Heaviside function with $\theta_{ii}=1/2$).  More precisely,
one must {\em rewrite\/} the operator valued Hamiltonian $H=-\beta^2
\slashed{D}^2/2$ in Weyl ordered form.  {\em Thereafter\/} one obtains
the Lagrangian\footnote{Note that the exponent of the path integral is
then $-(1/\beta)\int {\cal L}$ so that if one continues the Euclidean
time $t\rightarrow it$ back to real time the exponent becomes
$i(1/\beta)\int (p_\mu \dot x^\mu+(i/2)\psi_r\dot\psi^r-H)$, {\it
i.e.}~none other than the usual Legendre transformation relating
Lagrangian and Hamiltonian.}
\begin{equation}
{\cal L}=-ip_\mu \dot x^\mu+\frac{1}{2}\psi_r\dot\psi^r+H\, .
\end{equation}
by replacing operators by $c$-numbers in the Weyl ordered Hamiltonian.
In general one then finds additional vertices, over and above the classical 
result, of order $\beta$ and $\beta^2$. The exact propagators 
are then those given in~\cite{boer4,boer5} and in our conventions in~\cite{wald1}.
Note that employing the vertices of~\eqn{e:action} and propagators of
table~\ref{t:propagators} along with the above rule for products of distributions
even the overall normalisation of the path integral as given in~\eqn{e:pathrep}
is exactly correct.

Our task then is to rewrite the operator $H=-\beta^2\slashed{D}^2/2$
in Weyl ordered form, {\it i.e.}~in terms of symmetrised products of
the bose operators $x^\mu$ and $p_\mu$ and antisymmetrised products
for the fermions $\psi^r$.  In particular, for the fermions, this just
means that all products of Dirac matrices should be written in the
basis $\gamma^r,\gamma^{rs}, \gamma^{rst}$ and $\gamma^{rstu}$.  This
computation will involve three different spin connections so to avoid
confusion let us spell out our notations.  The Dirac operator is
\begin{equation}
\slashed{D}=
\slashed{D}(e,A)=
\gamma^\mu  D(\omega)_\mu\, , \quad \gamma^\mu=e_r{}^\mu \gamma^r\, ,
\quad D(\omega)_\mu=\partial_\mu+\tfrac{1}{4}\omega_{\mu rs}\gamma^{rs}\, .
\end{equation}
Here $D(\omega)_\mu$ denotes an operator depending on the spin
connection $\omega_{\mu rs}=\omega(e)_{\mu rs}+A_{\mu rs}$ where
$\omega(e)_{\mu rs}$ is the torsion free connection and $A_{\mu rs}$
is a completely antisymmetric contortion tensor.  The classical
($\beta$ independent) contribution to $H$ has been computed
in~\cite{mavr2} and yields the Hamiltonian for a spinning particle
depending on a third connection $\widehat\omega_{\mu rs}=\omega_{\mu
rs}+2A_{\mu rs}= \omega(e)_{\mu rs}+3A_{\mu rs}$ with contortion three
times the usual one.  Hence we introduce the operator
$D(\widehat\omega)_\mu$ identical to the previous definition except
depending on the new connection $\widehat{\omega}_{\mu rs}$. We
reserve the symbol $D_\mu(\equiv e_\mu{}^rD_r)$ for the torsion free
covariant derivative.

We now make some simple algebraic manipulations
\begin{align}
\label{e:Weyl/2}
\slashed{D}^2/2&=\frac{1}{2}(g^{\mu\nu} D(\omega)_\mu
   -\Gamma^\nu{}_\mu{}^\mu+\gamma^{\mu\nu} D(\omega)_\mu
   +A_{rs}{}^\nu\gamma^r\gamma^s) D(\omega)_\nu\\
  &= \frac{1}{2}\,g^{\mu\nu} ( D(\widehat\omega)_\mu D(\widehat\omega)_\nu 
     - \Gamma_{\mu\rho}{}^\rho D(\widehat\omega)_\nu  )\nonumber\\
  &\quad +\,\frac{1}{16}\, (R(\omega)_{\mu\nu rs}
                          - 2 A^\rho{}_{\mu\nu} A_{\rho rs}) 
                       \gamma^{\mu\nu}\gamma^{rs}
          +\frac{1}{4} (D_\mu A^\mu{}_{rs})\gamma^{rs} \,\label{e:harmony}
\end{align}
with $R_{\mu\nu rs}$ the curvature built from the spin connection
$\omega_{\mu rs}$ (our conventions for the Riemann tensor in the
presence of torsion are spelled out in the appendix).  The remainder
of the Weyl ordering computation for the first term in~\eqn{e:harmony}
follows the computation in~\cite{boer5} and the remaining terms
require only the identity
$\gamma^{\mu\nu}\gamma_{rs}=\gamma^{\mu\nu}{}_{rs}+4\,
\gamma^{[\mu}{}_{[s}e^{\nu]}{}_{r]} +2\,e^{[\mu}{}_{s}e^{\nu]}{}_{r}$.
Hence we find, making the identifications in~\eqn{e:identify}
\begin{equation}
\begin{aligned}
H&=\frac{1}{2}\, \Big[ g^{\mu\nu}
		      (p_\mu-\frac{i}{2}\widehat\omega_{\mu rs}\psi^r\psi^s)\,
		      (p_\nu-\frac{i}{2}\widehat\omega_{\nu tu}\psi^t\psi^u)
		   \Big]_W
    -\frac{1}{2}e^\mu{}_r (D_\mu
		      A_{stu})\Big[\psi^r\psi^s\psi^t\psi^u\Big]_W \\[.5em]
 &\quad -\frac{\beta^2}{8}\,(\Gamma^\sigma{}_{\mu\rho}\Gamma^{\rho\mu}{}_{\sigma}
+\frac{1}{2}\,\widehat{\omega}_{\mu rs}\widehat{\omega}^{\mu rs}
+3\,A_{rst}A^{rst})
\end{aligned}
\end{equation}
where $[\ldots]_W$ denotes Weyl ordering. We may now take the Weyl ordered
expression for $H$ and replace operators by $c$-numbers. Integrating
out the momenta $p_\mu$ one finds the Lagrangian 
\begin{equation}
\label{e:action}
\begin{aligned}
{\cal L}&=
\frac{1}{2}\,g_{\mu\nu}\dot x^\mu\,\dot x^\nu
+\frac{1}{2}\psi^r\Big(\delta_{rs}\frac{{\rm d}}{{\rm d}t}
+\dot x^\mu\widehat\omega_{\mu rs}\Big)\psi^s 
-\frac{1}{2}\, D_r A_{stu}\,\psi^r\psi^s\psi^t\psi^u \\[.5em]
&\quad -\frac{\beta^2}{8}\Big(\Gamma^\sigma{}_{\mu\rho}\Gamma^{\rho\mu}{}_{\sigma}
+\frac{1}{2}\widehat\omega_{\mu rs}\widehat\omega^{\mu rs}
 + 3\, A_{rst} A^{rst}\Big)\, .
\end{aligned}
\end{equation}
At this point the vertices of the theory are those defined
by~\eqn{e:action} and propagators along with products of distributions
are those found according to the midpoint rule.  Observe that the term
at order $\beta$ vanishes identically.  For zero torsion, the above
result has appeared in \cite{alfa1}.  Note also that the first two
$O(\beta^2)$ terms in the second line of~\eqn{e:action} are neither
general coordinate (GC) nor local Lorentz (LL) invariant.  {}From first
principles however, we know that the result for the index must enjoy
these invariances, since varying the background fields $e_\mu{}^r$ and
$\omega_{\mu rs}$ with respect to GC and LL transformations, the
operator $\slashed{D}^2$ varies into a commutator
$[\slashed{D}^2,G+L]$, ($G$ and $L$ are the generators of GC and LL
transformations) which vanishes under the trace as $[\gamma^5,G+L]=0=
[\gamma^5,\slashed{D}^2]$.  The resolution of this apparent dichotomy
is simply that the propagators, based on the midpoint rule obtained by
Weyl ordering, are also not GC or LL covariant. Nonetheless, as we
shall see, the combination of these two ``evils'' will yield a GC and
LL invariant result.

\subsection{Loop expansion of the path integral}

We are now ready to compute the path integral via a loopwise expansion
in the Planck constant $\beta$ to orders $\beta^{-1}$ and $\beta^0$
employing the Feynman rules for the finite-time path integral of
\cite{boer4,boer5}.  
The split into free and interacting parts is made
via the expansion around backgrounds $y^\mu$ and $\Psi$ so that
$x^\mu\rightarrow y^\mu+q^\mu$ and $\psi^r\rightarrow \Psi^r +\eta^r$.
The spin connection $\widehat{\omega}$ is now an
independent background field hence in our diagrammatic computation we
make a Riemann normal coordinate expansion for the metric only
(details can be found in the references just mentioned as well as
in~\cite{wald1}).  
We have listed the propagators in table~\ref{t:propagators} and some
of the required integrals in table~\ref{t:integrals}. Other useful
diagrammatic tricks and identities for this type of computation have
been developed in~\cite{wald1}.

We only need compute graphs quartic in background fermions $\Psi^r$
(in the diagrams they are denoted by a dash on the external line) in
order to saturate the Grassmann integration ${\rm d}^4\Psi$
in~\eqn{e:pathrep} which in turn yields  a four-dimensional
Levi-Civita symbol.  There are several ``tree level''
contributions\footnote{To avoid confusion, note that we only compute contributions
of order $\beta^{-1}$ and $\beta^0$, however, since we must compute the partition 
function $Z[y^\mu,\Psi^r]$, 
the generating function of {\em all}\/ diagrams, disconnected products of 
``two loop'' and ``tree'' graphs
will produce $\beta$ independent results which must not be neglected.} 
(those contributions arising from the terms in the
action that are independent of quantum fields). To order $\beta^0$,
\begin{multline}
\label{e:tree}
\Delta_{\text{tree}}(\slashed{D}) = \\[1em]
\int_{\cal M}\! {\rm d}^4y\, \frac{\sqrt{g}}{(2\pi i)^2}\, \epsilon^{rstu}\,
\bigg[\,
\frac{1}{2\beta}\, D_r A_{stu}\,
+\,\frac{1}{32}\,  D_r A_{stu}\, 
\widehat \omega_{\nu mn} \widehat \omega^{\nu mn}\,+\,
\frac{3}{16}\,A_{mnp}A^{mnp}\, D_r A_{stu}\,
\bigg] \, ,
\end{multline}
beginning with the leading $1/\beta$ contribution to which we draw the
reader's attention. This
contribution is proportional to the so-called Nieh--Yan tensor
\cite{nieh1,nieh2}. As expressed above it is clear that this
tensor vanishes for exact torsion; alternatively one can re-express the
Nieh-Yan contribution into a perhaps more familiar form
by writing
\begin{equation}
\label{e:FvsNY}
2\, \partial_{[\mu}A_{\nu\rho\sigma]} = R(\omega)_{[\mu\nu\rho\sigma]} - 2\, A_{\lambda[\mu\nu}  
A^\lambda{}_{\rho\sigma]} \,  .
\end{equation}
where $R(\omega)_{\mu\nu\rho\sigma}$ is the Riemann curvature of the
torsion-full connection $\omega_{\mu rs}=\omega(e)_{\mu rs}+A_{
\mu rs}$.
We will discuss this contribution, which has led to some confusion in
the literature, in more detail in the conclusion. Observe that the
order $\beta^0$ terms are also proportional to the curl of the
torsion, though not immediately recognisable as a topological
invariant as they are only a part of the result at this order in the
Planck constant. Furthermore, observe that the product of the
Nieh--Yan tree level tree level $\beta^{-1}$ contribution and the
Christ and Lee Weyl ordering counter terms yield the second and third
${\cal O}(\beta^0)$ terms in~\eqn{e:tree}. This provides the long sought after
explicit confirmation that these two loop counter terms (they appear
with a factor $\beta^2$ in the action~\eqn{e:action}) are actually
necessary in a rigorous definition of the path integral.

The remaining graphs involve one- and two-loop integrals and are given
in tables~\ref{t:oneloop} and \ref{t:twoloop}. Note that we have not
drawn graphs obtained by different locations of the dots, which denote
time derivatives on propagators, although these contributions are, of
course, included in the quoted results.  Adding all these
contributions together our result for the regulated trace of
$\gamma_5$ is
\begin{multline}
\label{e:indexwithtorsion}
\Delta(\slashed{D})=
\int_{\cal M}\! {\rm d}^4y\, \frac{\sqrt{g}}{(2\pi i)^2}\,
  \epsilon^{rstu} \, \bigg\{
      \frac{1}{2\beta} \, D_r A_{stu}\,
 +\,\frac{1}{192}\Big[
       R(\widehat \omega)_{\mu\nu rs}R(\widehat \omega)^{\mu\nu}{}_{tu}
\\[1em] 
     + 8\, D^\mu D_\mu  D_r  A_{stu}
     + 36\, A_{mnp}A^{mnp}\, D_r A_{stu} 
     + 4\, R\, D_r A_{stu} \Big]
     \bigg\} \, + \, {\cal O}(\beta)\, .
\end{multline}
Note that the contracted indices in the term quadratic in the Riemann
tensors are {\em curved}\/ ones since in the presence of torsion
the $R(\widehat\omega)_{\mu\nu rs}\neq R(\widehat\omega)_{rs\mu\nu }$
(see~\eqn{e:RRptors}).
The graphs for this term are not displayed in
the table but can be easily computed along the lines of~\cite{wald1}.
Also, observe that the scalar curvature on the second line of~\eqn{e:indexwithtorsion}
is that of the torsion-free Riemann curvature since it comes from the figure
eight graph on the last line of table~\ref{t:twoloop} in which one expands the
metric $g_{\mu\nu}(y+q)$ to second order in quantum fluctuations $q(t)^\mu$ which
in Riemann normal coordinates yields the torsion free Riemann curvature.

A useful test of the result \eqn{e:indexwithtorsion} is
independence of the choice of complexification of the original
Majorana spinors required to employ a coherent state formalism for the
fermions. The three terms from the one-loop diagrams and the single
term from the two-loop diagram do indeed vanish upon using the
Schouten identity, by virtue of the fact that $K^{mn}$ is
anti-symmetric.

To facilitate comparison with existing results and see that the result
is a total derivative it is useful to extract the dependence on the
torsion free curvature which we denote by $R_{\mu\nu\rho}{}^\sigma$
where $R_{\mu\nu\rho}{}^\sigma V_\sigma=[D_\mu,D_\nu]V_\rho$.
Furthermore, we rewrite the result in terms of the axial contortion
vector $A_r=\epsilon_{rstu}A^{stu}$.  Using \eqn{e:RRptors} one
derives that
\begin{equation}
\begin{aligned}
R(\widehat\omega)_{\mu\nu rs} 
R(\widehat\omega)^{\mu\nu}{}_{tu}\, 
  \epsilon^{rstu} 
&= \Big[ R_{\mu\nu\rho\sigma} R_{\kappa\lambda}{}^{\rho\sigma} 
+ 2\, D_\mu  A_\nu D_\kappa  A_\lambda \Big] \,\frac{1}{\sqrt{g}}\,\epsilon^{\mu\nu\kappa\lambda} \\[1em]
&\quad - 4 \Big[ D^\mu  A_\mu  A^\nu  A_\nu 
              - D_\mu  A_\nu  A^\mu  A^\nu \Big]
\, -\, 8\, D_\mu  A_\nu R^{\mu\nu} \, .
\end{aligned}
\end{equation}
The cross terms between $R$ and the torsion vanish identically by
virtue of the Ricci symmetry $R_{\mu[\nu\rho\sigma]}=0$ and so do the
terms quartic in the torsion. The second line of
\eqn{e:indexwithtorsion} is easily written as a covariant derivative
by commuting the triple derivatives using
\begin{equation}
D^\mu D_\mu  D_\nu  A^\nu 
 = D_\nu D^\mu D_\mu A^\nu + D_{\mu}A_\nu R^{\mu\nu}
+ \tfrac{1}{2}  A^\mu D_\mu R  \, .
\end{equation}
This brings \eqn{e:indexwithtorsion} into a form derived previously by
Obukhov~\cite{obuk2} using heat kernel methods\footnote{Obukhov's work
has been re-examined by Yajima and Kimura~\cite{yaji1,yaji2} who found,
again in a heat kernel approach, the same result for the bulk index
with torsion but disagree upon the precise details of his derivation
of the heat kernel coefficients and claim that it leads to incorrect
results for the Lorentz anomaly in the presence of torsion. We do not
wish to enter this debate here and it suffices to say that the path
integral techniques presented in this paper can equally well be
employed for other anomalies and could, therefore, provide an
independent resolution of this issue.},
\begin{multline}
\label{e:bulkresult}
\Delta(\slashed{D})=
\int_{\cal M}\! {\rm d}^4y\, \frac{\sqrt{g}}{(2\pi i)^2}\,
 \bigg\{
\frac{1}{2\beta}D_\mu A^\mu +\frac{1}{192}
\frac{1}{\sqrt{g}}\,\epsilon^{\rho\sigma\kappa\lambda} \Big[ 
R_{\mu\nu \rho\sigma} R^{\mu\nu}{}_{\kappa\lambda}
+ \frac{1}{2}  F_{\rho\sigma}  F_{\kappa\lambda} \Big]
+ \frac{1}{24}\, D_\mu {\cal K}^\mu \bigg\}\,\\[.5em]
+{\cal O}(\beta) \, .
\end{multline}
with
\begin{equation}
\label{e:no_isaacs_dog_was_diamond}
{\cal K}^\mu = \Big( D^{\nu} D_{\nu} + \tfrac{1}{4}  A^\nu  A_\nu + \tfrac{1}{2} R\Big)\, 
 A^\mu \quad , \quad
F_{\mu\nu}=2D_{[\mu} A_{\nu]}\, .
\end{equation}

In the case that the Nieh-Yan tensor vanishes, which is manifestly so
when the torsion is exact, the above expression
\eqn{e:indexwithtorsion} reduces to the term quadratic in
$R(\widehat\omega)$ which coincides with the result derived by Mavromatos
some time ago \cite{mavr2}. 

As long as the ${\cal K}^\mu$ vector is smooth (as it is for smooth
metric and torsion), the additional contribution $D_\mu {\cal K}^\mu$
to the bulk index in the presence of torsion will only receive
contributions from the boundary. The subtle interplay
between the three order $\beta^0$ terms in \eqn{e:bulkresult} can be
clearly seen in the explicit example given in subsection~\ref{ss:bulk_example} below.

We stress that the Nieh-Yan term in the action \eqn{e:action} 
along with the Christ and Lee two-loop Weyl ordering counter terms
have been
crucial in obtaining this result (see the two-loop diagrams in
table~\ref{t:twoloop} as well as~\eqn{e:tree}).  Although
the $1/\beta$ Nieh--Yan bulk contribution to index has appeared in the
literature before~\cite{obuk1,soo1,miel1,chan4,chan5,chan6}, in our path
integral approach one sees clearly
its importance
for the order $\beta^0$ terms and hence the necessity to include it
when computing the bulk index in the presence of torsion. We will
comment
further on this term in the conclusion.

Finally, we observe that the folkloric statement; ``the chiral anomaly
is topological and therefore insensitive to the subtleties of the
precise definition of the path integral'' can be made very definitive
in the approach employed above. The reason being that in the absence
of torsion the different counter terms corresponding to varying
ordering schemes have no affect on the $\beta^0$ result since they are
all higher loop effects. Only once the torsion is turned on, so that
the Lagrangian includes the $1/\beta$ Nieh--Yan contribution must one
worry about the subtleties of ordering schemes etc.\ when computing
the $\beta$ independent anomaly.

\begin{fmffile}{torsionindex}

\fmfcmd{%
 style_def tight_wiggly expr p =
  save wiggly_len, wiggly_slope;
  wiggly_len = 2mm; wiggly_slope = 90;
  draw_photon p
 enddef;
 style_def tight_fermion expr p =
  save arrow_ang;
  arrow_ang = 10;
  draw_fermion p
 enddef;
}

\fmfcmd{%
 vardef middir(expr p,ang) =
   dir(angle direction length(p)/2 of p+ang)
 enddef;
 vardef bar (expr p, len, ang) =
   ((-len/2,0)--(len/2,0))
      rotated (ang + angle direction length(p)/2 of p)
      shifted point length(p) of p
 enddef;
 vardef dot (expr p, len, pos, dist) =
   ((-len/4,0)..(0,-len/4)..(len/4,0)..(0,len/4)..cycle)
      shifted point length(p)*pos of p
      shifted (dist*middir(p,-90))
 enddef;
 vardef hshift_dot (expr p, len, pos, dist) =
   ((-len/4,0)..(0,-len/4)..(len/4,0)..(0,len/4)..cycle)
      shifted point length(p)*pos of p
      shifted (dist,0)
 enddef;
 style_def single_dotted expr p =
     save wiggly_len, wiggly_slope;
     wiggly_len = 2mm; wiggly_slope = 90;
     draw (wiggly p);
     cfill dot(p,2mm,.2,3thick)
 enddef;
 style_def hshift_single_dotted expr p =
     save wiggly_len, wiggly_slope;
     wiggly_len = 2mm; wiggly_slope = 90;
     draw (wiggly p);
     cfill hshift_dot(p,2mm,.2,-3thick)
 enddef;
 style_def single_dotted_rev expr p =
     save wiggly_len, wiggly_slope;
     wiggly_len = 2mm; wiggly_slope = 90;
     draw (wiggly p);
     cfill dot(p,2mm,.8,3thick)
 enddef;
 style_def double_dotted expr p =
     save wiggly_len, wiggly_slope;
     wiggly_len = 2mm; wiggly_slope = 90;
     draw (wiggly p);
     cfill dot(p,2mm,.2,3thick);
     cfill dot(p,2mm,.8,3thick)
 enddef;
 style_def bg_fermion expr p =
     cdraw p;
     ccutdraw bar (p, 2mm, 90)
 enddef;
}

\begin{table}
\begin{equation*}
\begin{aligned}
\vcenter{\vbox{\hbox{
\begin{fmfgraph*}(70,30)
\fmfpen{.5pt}
\fmfleft{cl}
\fmfright{cr}
\fmf{tight_wiggly}{cl,cr}
\fmffreeze
\fmfv{label.angle=90,label.dist=6pt,label=$\scriptstyle \mu$}{cl}
\fmfv{label.angle=90,label.dist=6pt,label=$\scriptstyle \nu$}{cr}
\fmfv{d.sh=circle,d.siz=3thin}{cl}
\fmfv{d.sh=circle,d.siz=3thin}{cr}
\end{fmfgraph*}
}}} \;\; &= \; \langle\, q(s)^\mu q(t)^\nu\,\rangle && = g^{\mu\nu} \Big[
t(1-s)\theta(s-t)+s(1-t)\theta(t-s)\Big]\, ,\\
\vcenter{\vbox{\hbox{
\begin{fmfgraph*}(70,30)
\fmfpen{.5pt}
\fmfleft{cl}
\fmfright{cr}
\fmf{tight_fermion}{cl,cr}
\fmffreeze
\fmfv{label.angle=90,label.dist=6pt,label=$\scriptstyle m$}{cl}
\fmfv{label.angle=90,label.dist=6pt,label=$\scriptstyle n$}{cr}
\fmfv{d.sh=circle,d.siz=3thin}{cl}
\fmfv{d.sh=circle,d.siz=3thin}{cr}
\end{fmfgraph*}
}}} \;\; &= \; \langle\, \eta(s)^m \eta(t)^n\,\rangle && =
\tfrac{1}{2}\eta^{mn}\Big[ \theta(s-t)-\theta(t-s)\Big] + K^{mn}\, .\\
\end{aligned}
\end{equation*}
\caption{Propagators for the action \eqn{e:action}. The symbol
$K^{mn}$ parametrises the choice of complexification of the Majorana
spinors, as explained in \cite{boer5}.}
\label{t:propagators}
\end{table}


\begin{table}
\begin{equation*}
\begin{aligned}
\vcenter{\vbox{\hbox{
\begin{fmfgraph*}(80,30)
\fmfpen{.5pt}
\fmfleft{cl}
\fmfright{cr}
\fmffixed{(0,0)}{cl,ccl}
\fmffixed{(0,0)}{cr,ccr}
\fmffixed{(-6thin,0)}{m,mml}
\fmffixed{(6thin,0)}{m,mmr}
\fmf{tight_fermion}{cl,m}
\fmf{tight_fermion}{cr,m}
\fmffreeze
\fmfv{label.angle=90,label.dist=6pt,label=$\scriptstyle n$}{mml}
\fmfv{label.angle=90,label.dist=6pt,label=$\scriptstyle p$}{mmr}
\fmfv{label.angle=90,label.dist=6pt,label=$\scriptstyle m$}{cl}
\fmfv{label.angle=90,label.dist=6pt,label=$\scriptstyle q$}{cr}
\fmfv{label.angle=-90,label.dist=6pt,label=$\scriptstyle s$}{ccl}
\fmfv{label.angle=-90,label.dist=6pt,label=$\scriptstyle t$}{ccr}
\fmfv{d.sh=circle,d.siz=3thin}{cl}
\fmfv{d.sh=cross,d.siz=6thin}{m}
\fmfv{d.sh=circle,d.siz=3thin}{cr}
\end{fmfgraph*}
}}} \;\; &= \;\; \delta^{mn}\delta^{pq}\big[ \tfrac{1}{2}(s-t)\,  
\epsilon(s-t) - \tfrac{1}{4}\big]\\
&\quad\quad+ (s-\tfrac{1}{2})\, \delta^{mn} K^{qp} +  
(t-\tfrac{1}{2})\, \delta^{qp}K^{mn} + K^{mn}K^{qp}\, ,\\
\vcenter{\vbox{\hbox{
\begin{fmfgraph*}(80,30)
\fmfpen{.5pt}
\fmfleft{cl}
\fmfright{cr}
\fmf{tight_fermion}{cl,m}
\fmf{tight_fermion}{cr,m}
\fmffixed{(0,0)}{cr,ccr}
\fmffixed{(-6thin,0)}{m,mml}
\fmffixed{(6thin,0)}{m,mmr}
\fmffreeze
\fmfv{label.angle=90,label.dist=6pt,label=$\scriptstyle m$}{cl}
\fmfv{label.angle=90,label.dist=6pt,label=$\scriptstyle n$}{mml}
\fmfv{label.angle=90,label.dist=6pt,label=$\scriptstyle p$}{mmr}
\fmfv{label.angle=90,label.dist=6pt,label=$\scriptstyle q$}{cr}
\fmfv{label.angle=-90,label.dist=6pt,label=$\scriptstyle t$}{ccr}
\fmfv{d.sh=cross,d.siz=6thin}{cl}
\fmfv{d.sh=cross,d.siz=6thin}{m}
\fmfv{d.sh=circle,d.siz=3thin}{cr}
\end{fmfgraph*}
}}} \;\; &= \;\;\tfrac{1}{2}(t^2-t)\delta^{mn}\delta^{pq} -  
(t-\tfrac{1}{2})\delta^{pq}K^{mn} + K^{mn}K^{qp}\, ,\\
\vcenter{\vbox{\hbox{
\begin{fmfgraph*}(80,30)
\fmfpen{.5pt}
\fmfleft{cl}
\fmfright{cr}
\fmf{single_dotted}{m,cl}
\fmf{tight_fermion}{m,cr}
\fmffixed{(0,0)}{cl,ccl}
\fmffixed{(0,0)}{cr,ccr}
\fmffreeze
\fmfv{label.angle=90,label.dist=6pt,label=$\scriptstyle m$}{m}
\fmfv{label.angle=90,label.dist=6pt,label=$\scriptstyle n$}{cr}
\fmfv{label.angle=-90,label.dist=6pt,label=$\scriptstyle s$}{ccl}
\fmfv{label.angle=-90,label.dist=6pt,label=$\scriptstyle t$}{ccr}
\fmfv{d.sh=circle,d.siz=3thin}{cl}
\fmfv{d.sh=cross,d.siz=6thin}{m}
\fmfv{d.sh=circle,d.siz=3thin}{cr}
\end{fmfgraph*}
}}} \;\; &= \;\; -\;\;
\vcenter{\vbox{\hbox{
\begin{fmfgraph*}(50,30)
\fmfpen{.5pt}
\fmfleft{cl}
\fmfright{cr}
\fmf{tight_wiggly}{cl,cr}
\fmffixed{(0,0)}{cl,ccl}
\fmffixed{(0,0)}{cr,ccr}
\fmffreeze
\fmfv{label.angle=90,label.dist=6pt,label=$\scriptstyle m$}{cl}
\fmfv{label.angle=90,label.dist=6pt,label=$\scriptstyle n$}{cr}
\fmfv{label.angle=-90,label.dist=6pt,label=$\scriptstyle s$}{ccl}
\fmfv{label.angle=-90,label.dist=6pt,label=$\scriptstyle t$}{ccr}
\fmfv{d.sh=circle,d.siz=3thin}{cl}
\fmfv{d.sh=circle,d.siz=3thin}{cr}
\end{fmfgraph*}
}}} \, .
\end{aligned}
\end{equation*}
\begin{equation*}
\begin{aligned}
\vcenter{\vbox{\hbox{
\begin{fmfgraph*}(70,45)
\fmfpen{.5pt}
\fmfleft{cl}
\fmfright{cr}
\fmffixed{(0,0)}{cl,ccl}
\fmffixed{(0,0)}{cr,ccr}
\fmf{tight_fermion,left=.6}{cl,cr}
\fmf{tight_fermion,right=.6}{cl,cr}
\fmffreeze
\fmfv{label.angle=90,label.dist=6pt,label=$\scriptstyle m$}{cl}
\fmfv{label.angle=90,label.dist=6pt,label=$\scriptstyle n$}{cr}
\fmfv{label.angle=-90,label.dist=6pt,label=$\scriptstyle p$}{ccl}
\fmfv{label.angle=-90,label.dist=6pt,label=$\scriptstyle q$}{ccr}
\fmfv{d.sh=cross,d.siz=6thin}{cl}
\fmfv{d.sh=cross,d.siz=6thin}{cr}
\end{fmfgraph*}
}}} \;\; &= \;\; \tfrac{1}{4}\delta^{mn}\delta^{pq} + K^{mn} K^{pq}\,
, &
\vcenter{\vbox{\hbox{
\begin{fmfgraph*}(70,45)
\fmfpen{.5pt}
\fmfleft{cl}
\fmfright{cr}
\fmf{tight_fermion,left=.5}{cl,cr}
\fmf{double_dotted,right=.5}{cl,cr}
\fmffreeze
\fmfv{l.angle=90,label.dist=6pt,label=$\scriptstyle n$}{cl}
\fmfv{l.angle=90,label.dist=6pt,label=$\scriptstyle m$}{cr}
\fmfv{d.sh=cross,d.siz=6thin}{cl}
\fmfv{d.sh=cross,d.siz=6thin}{cr}
\end{fmfgraph*}
}}} \;\; &= \;\;0\, ,\\
\vcenter{\vbox{\hbox{
\begin{fmfgraph*}(70,45)
\fmfpen{.5pt}
\fmfleft{cl}
\fmfright{cr}
\fmf{tight_fermion,left=.5}{cl,cr}
\fmf{single_dotted,right=.5}{cl,cr}
\fmffreeze
\fmfv{l.angle=90,label.dist=6pt,label=$\scriptstyle n$}{cl}
\fmfv{l.angle=90,label.dist=6pt,label=$\scriptstyle m$}{cr}
\fmfv{d.sh=cross,d.siz=6thin}{cl}
\fmfv{d.sh=cross,d.siz=6thin}{cr}
\end{fmfgraph*}
}}} \;\; &= \;\;-\tfrac{1}{6}\delta^{nm} \, ,&\quad
\vcenter{\vbox{\hbox{
\begin{fmfgraph*}(80,30)
\fmfpen{.5pt}
\fmfleft{cl}
\fmfright{cr}
\fmf{tight_fermion}{cl,m}
\fmf{tight_fermion}{cr,m}
\fmffixed{(-6thin,0)}{m,mml}
\fmffixed{(6thin,0)}{m,mmr}
\fmffreeze
\fmfv{label.angle=90,label.dist=6pt,label=$\scriptstyle m$}{cl}
\fmfv{label.angle=90,label.dist=6pt,label=$\scriptstyle n$}{mml}
\fmfv{label.angle=90,label.dist=6pt,label=$\scriptstyle p$}{mmr}
\fmfv{label.angle=90,label.dist=6pt,label=$\scriptstyle q$}{cr}
\fmfv{d.sh=cross,d.siz=6thin}{cl}
\fmfv{d.sh=cross,d.siz=6thin}{m}
\fmfv{d.sh=cross,d.siz=6thin}{cr}
\end{fmfgraph*}
}}} \;\; &= \;\;-\tfrac{1}{12}\delta^{mn}\delta^{pq} + K^{mn}  
K^{qp} \, , \\
\vcenter{\vbox{\hbox{
\begin{fmfgraph*}(70,45)
\fmfpen{.5pt}
\fmfleft{cl}
\fmfright{cr}
\fmfipair{crb}
\fmf{tight_fermion}{cl,m}
\fmf{phantom,tension=3}{m,cr}
\fmffreeze
\fmfiequ{ypart(crb)}{ypart(se)}
\fmfiequ{xpart(crb)}{xpart(vloc(__m))}
\fmffreeze
\fmfi{hshift_single_dotted}{vloc(__m){sw-ne} .. {right}crb.. {nw-se}vloc(__m)}
\fmfv{l.angle=90,label.dist=6pt,label=$\scriptstyle n$}{cl}
\fmfv{l.angle=90,label.dist=6pt,label=$\scriptstyle m$}{m}
\fmfv{d.sh=cross,d.siz=6thin}{cl}
\fmfv{d.sh=cross,d.siz=6thin}{m}
\end{fmfgraph*}
}}} \;\; &= \;\;\delta^{nm}\tfrac{1}{12} \, ,&
%
%
\vcenter{\vbox{\hbox{
\begin{fmfgraph*}(50,50)
\fmfpen{.5pt}
\fmftop{tl,tr}
\fmfbottom{cb}
\fmf{tight_fermion}{cb,tl}
\fmf{tight_fermion}{cb,tr}
\fmf{double_dotted}{tl,tr}
\fmffixed{(-6thin,0)}{cb,cbl}
\fmffixed{(6thin,0)}{cb,cbr}
\fmffreeze
\fmfv{label.dist=6pt,label=$\scriptstyle n$}{tl}
\fmfv{label.dist=6pt,label=$\scriptstyle m$}{tr}
\fmfv{label.dist=2pt,label=$\scriptstyle p$}{cbl}
\fmfv{label.dist=2pt,label=$\scriptstyle q$}{cbr}
\fmfv{d.sh=cross,d.siz=6thin}{tl}
\fmfv{d.sh=cross,d.siz=6thin}{tr}
\fmfv{d.sh=cross,d.siz=6thin}{cb}
\end{fmfgraph*}
}}} \;\;\; &= \;\;-\tfrac{1}{6} \delta^{nm}\delta^{pq}\, , \\[1em]
\vcenter{\vbox{\hbox{
\begin{fmfgraph*}(80,50)
\fmfpen{.5pt}
\fmftop{cl}
\fmfbottom{cr}
\fmf{phantom}{cl,m,cr}
\fmf{tight_fermion,left=90,tension=.5}{m,m}
\fmf{tight_fermion,tension=.5}{m,m}
\fmffreeze
\fmffixed{(0,0)}{m,m2}
\fmffixed{(0,0)}{m,m3}
\fmffixed{(0,0)}{m,m4}
\fmfv{l.dist=12pt,l.angle=120,l=$\scriptstyle m$}{m}
\fmfv{l.dist=12pt,l.angle=60,l=$\scriptstyle n$}{m2}
\fmfv{l.dist=12pt,l.angle=-120,l=$\scriptstyle p$}{m3}
\fmfv{l.dist=12pt,l.angle=-60,l=$\scriptstyle q$}{m4}
\fmfv{d.sh=cross,d.siz=6thin}{m}
\end{fmfgraph*}
}}} \; &= \;\; K^{mp} K^{qn}\, ,&
\vcenter{\vbox{\hbox{
\begin{fmfgraph*}(80,50)
\fmfpen{.5pt}
\fmftop{ct}
\fmfbottom{cb}
\fmf{tight_fermion,left=.9}{ct,cb}
\fmf{tight_fermion,right=.9}{ct,cb}
\fmffreeze
\fmf{double_dotted}{ct,cb}
\fmffixed{(0,0)}{ct,ct2}
\fmffixed{(0,0)}{cb,cb2}
\fmfv{l.dist=12pt,l.angle=180,l=$\scriptstyle m$}{ct}
\fmfv{l.dist=12pt,l.angle=180,l=$\scriptstyle p$}{cb}
\fmfv{l.dist=12pt,l.angle=0,l=$\scriptstyle n$}{ct2}
\fmfv{l.dist=12pt,l.angle=0,l=$\scriptstyle q$}{cb2}
\fmfv{d.sh=cross,d.siz=6thin}{ct}
\fmfv{d.sh=cross,d.siz=6thin}{cb}
\end{fmfgraph*}
}}} \; &= \;\; -\tfrac{1}{4}\delta^{mp}\delta^{nq}\, ,\\
\vcenter{\vbox{\hbox{
\begin{fmfgraph*}(70,45)
\fmfpen{.5pt}
\fmfleft{cl}
\fmfright{cr}
\fmfipair{crb}
\fmf{tight_fermion}{cl,m}
\fmf{phantom,tension=3}{m,cr}
\fmffreeze
\fmffixed{(0,0)}{m,m2}
\fmffixed{(0,0)}{m,m3}
\fmfiequ{ypart(crb)}{ypart(se)}
\fmfiequ{xpart(crb)}{xpart(vloc(__m))}
\fmffreeze
\fmfi{tight_fermion}{vloc(__m){sw-ne} .. {right}crb.. {nw-se}vloc(__m)}
\fmfv{l.angle=90,label.dist=6pt,label=$\scriptstyle n$}{cl}
\fmfv{l.angle=90,label.dist=6pt,label=$\scriptstyle m$}{m}
\fmfv{l.angle=-150,label.dist=10pt,label=$\scriptstyle p$}{m2}
\fmfv{l.angle=-30,label.dist=10pt,label=$\scriptstyle q$}{m3}
\fmfv{d.sh=cross,d.siz=6thin}{cl}
\fmfv{d.sh=cross,d.siz=6thin}{m}
\end{fmfgraph*}
}}} \;\; &= \;\; K^{nm} K^{pq}  \, . &
%
%
\vcenter{\vbox{\hbox{
\begin{fmfgraph*}(70,30)
\fmfpen{.5pt}
\fmfleft{cl}
\fmfright{cr}
\fmf{tight_fermion}{cl,cr}
\fmffreeze
\fmfv{label.angle=90,label.dist=6pt,label=$\scriptstyle r$}{cl}
\fmfv{label.angle=90,label.dist=6pt,label=$\scriptstyle s$}{cr}
\fmfv{d.sh=cross,d.siz=6thin}{cl}
\fmfv{d.sh=cross,d.siz=6thin}{cr}
\end{fmfgraph*}
}}} \;\; &= \; 0 \, , \\
\end{aligned}
\end{equation*}
\caption{A number of useful integrals obtained by straightforward
application of the propagators in table~\ref{t:propagators} where the
definition of $K^{mn}$ may also be found. The remaining
integrals required for our computation as well as a detailed
explanation of the diagrammatic notation above can be found in \cite{wald1}.}
\label{t:integrals}
\end{table}


\begin{table}
\begin{equation*}
\begin{aligned}
\vcenter{\vbox{\hbox{
\begin{fmfgraph}(60,60)
\fmfpen{.5pt}
\fmfleft{tl,bl}
\fmfright{tr,br}
\fmf{bg_fermion}{ml,tl}
\fmf{bg_fermion}{mr,tr}
\fmf{bg_fermion}{ml,bl}
\fmf{bg_fermion}{mr,br}
\fmf{tight_fermion,tension=.3,left=.7}{ml,mr,ml}
\fmffreeze
\end{fmfgraph}
}}} \;\; &= \;\; -\frac{1}{64} \, \Psi^{rstu}\,  
\widehat F_{rsnm}\widehat F_{tu}{}^{mn} + \frac{1}{16} \Psi^{rstu}
  \widehat F_{rsnm} \widehat F_{tupq} K^{mp} K^{nq} \, ,\\
 & \\
\vcenter{\vbox{\hbox{
\begin{fmfgraph}(60,60)
\fmfpen{.5pt}
\fmfleft{tl,bl}
\fmfright{tr,br}
\fmf{bg_fermion}{m,tl}
\fmf{bg_fermion}{m,tr}
\fmf{bg_fermion}{m,bl}
\fmf{bg_fermion}{m,br}
\fmffreeze
\fmf{tight_wiggly}{m,m}
\end{fmfgraph}
}}} \;\; &= \;\; -\frac{1}{288} \, \Psi^{rstu}\,  
\partial^\mu\partial_\mu \widehat F_{rstu}\, ,\\
 & \\
\vcenter{\vbox{\hbox{
\begin{fmfgraph}(60,60)
\fmfpen{.5pt}
\fmfleft{tl,cl,bl}
\fmfright{cr}
\fmf{bg_fermion}{ml,tl}
\fmf{bg_fermion}{ml,bl}
\fmf{bg_fermion}{mr2,cr}
\fmf{tight_fermion,tension=.3}{mr2,mr,ml}
\fmffreeze
\fmf{double_dotted,right=.9}{mr,mr2}
\fmf{bg_fermion}{ml,cl}
\end{fmfgraph}
}}} \;\; &= \;\; \frac{2}{288} \Psi^{rstu} \widehat F_{rstn}  
\widehat\omega^{\mu nm} \widehat\omega_{\mu um} \, ,\\
 & \\
\vcenter{\vbox{\hbox{
\begin{fmfgraph}(60,60)
\fmfpen{.5pt}
\fmfleft{tl,cl,bl}
\fmfright{cr}
\fmfipair{crt}
\fmf{bg_fermion}{ml,tl}
\fmf{bg_fermion}{ml,bl}
\fmf{bg_fermion}{mr,cr}
\fmf{tight_fermion,tension=.3}{mr,ml}
\fmffreeze
\fmfiequ{ypart(crt)}{ypart(.8[se,ne])}
\fmfiequ{xpart(crt)}{xpart(vloc(__mr))}
\fmffreeze
\fmfi{tight_fermion}{vloc(__mr){left} .. {right}crt.. {left}vloc(__mr)}
\fmf{bg_fermion}{ml,cl}
\end{fmfgraph}
}}} \;\; &= \;\; -\frac{24}{288} \Psi^{rstu} \widehat F_{rstn}  
\widehat F_{upqm} K^{np} K^{qm} \, ,\\
 & \\
%
%
\vcenter{\vbox{\hbox{
\begin{fmfgraph}(60,60)
\fmfpen{.5pt}
\fmfleft{tl,bl}
\fmfright{tr,br}
\fmf{tight_fermion}{m,ml}
\fmf{tight_fermion}{m,mr}
\fmf{bg_fermion}{ml,tl}
\fmf{bg_fermion}{mr,tr}
\fmf{bg_fermion}{m,bl}
\fmf{bg_fermion}{m,br}
\fmffreeze
\fmf{double_dotted}{ml,mr}
\end{fmfgraph}
}}} \;\; &= \;\; -\frac{12}{288} \Psi^{rstu} \widehat F_{rs}{}^{nm}  
\widehat\omega_{\mu tm} \widehat\omega^\mu{}_{un}\, ,\\
\end{aligned}
\end{equation*}
\begin{equation*}
\begin{aligned}
\left(\vcenter{\vbox{\hbox{
\begin{fmfgraph}(40,40)
\fmfpen{.5pt}
\fmfleft{tl,bl}
\fmfright{tr,br}
\fmfipair{mr}
\fmf{bg_fermion}{m,tl}
\fmf{bg_fermion}{m,bl}
\fmf{phantom}{m,tr}
\fmf{phantom}{m,br}
\fmffreeze
\fmfiequ{ypart(mr)}{ypart(vloc(__m))}
\fmfiequ{xpart(mr)}{xpart(vloc(__tr))}
\fmffreeze
\fmfi{tight_fermion}{vloc(__m){up} .. {down}mr .. {up}vloc(__m)}
\end{fmfgraph}
}}}+
\vcenter{\vbox{\hbox{
\begin{fmfgraph}(40,40)
\fmfpen{.5pt}
\fmfleft{tl,bl}
\fmfright{tr,br}
\fmfipair{mr}
\fmf{bg_fermion}{m,tl}
\fmf{bg_fermion}{m,bl}
\fmf{phantom}{m,tr}
\fmf{phantom}{m,br}
\fmffreeze
\fmfiequ{ypart(mr)}{ypart(vloc(__m))}
\fmfiequ{xpart(mr)}{xpart(vloc(__tr))}
\fmffreeze
\fmfi{single_dotted}{vloc(__m){up} .. {down}mr .. {up}vloc(__m)}
\end{fmfgraph}
}}}+
\vcenter{\vbox{\hbox{
\begin{fmfgraph}(40,40)
\fmfpen{.5pt}
\fmfleft{ll}
\fmfright{rr}
\fmf{bg_fermion}{ml,ll}
\fmf{bg_fermion}{mr,rr}
\fmf{tight_fermion,tension=.5}{ml,mr}
\fmffreeze
\fmf{double_dotted,right=.9}{mr,ml}
\end{fmfgraph}
}}}
\right)^2 \;\; &= \;\; \frac{1}{32} \Psi^{rstu} \widehat F_{rsmn}  
K^{mn} \widehat F_{tupq} K^{pq} \, ,\\
\end{aligned}
\end{equation*}
\caption{One-loop diagrams contributing to the path integral at  
order $\beta^0$. Only those graphs that are not already
zero because of symmetry properties of the background fields are
displayed. Note that we denote $\Psi^{rstu}\equiv\Psi^r\Psi^s\Psi^t\Psi^u$.}
\label{t:oneloop}
\end{table}

\begin{table}
\begin{equation*}
\begin{aligned}
\vcenter{\vbox{\hbox{
\begin{fmfgraph}(30,30)
\fmfpen{.5pt}
\fmfleft{tl,bl}
\fmfright{tr,br}
\fmf{bg_fermion}{m,tl}
\fmf{bg_fermion}{m,bl}
\fmf{bg_fermion}{m,tr}
\fmf{bg_fermion}{m,br}
\end{fmfgraph}
}}}
\vcenter{\vbox{\hbox{
\begin{fmfgraph}(80,50)
\fmfpen{.5pt}
\fmftop{ct}
\fmfbottom{cb}
\fmf{tight_fermion,left=.9}{ct,cb}
\fmf{tight_fermion,right=.9}{ct,cb}
\fmffreeze
\fmf{double_dotted}{ct,cb}
\fmffixed{(0,0)}{ct,ct2}
\fmffixed{(0,0)}{cb,cb2}
\end{fmfgraph}
}}} \; &= \;\; -\frac{(3/4)}{288}\widehat\omega_{\mu mn}\widehat\omega^{\mu  
mn} \Psi^{rstu} \widehat F_{rstu}\, ,\\
 & \\
\vcenter{\vbox{\hbox{
\begin{fmfgraph}(30,30)
\fmfpen{.5pt}
\fmfleft{tl,bl}
\fmfright{tr,br}
\fmf{bg_fermion}{m,tl}
\fmf{bg_fermion}{m,bl}
\fmf{bg_fermion}{m,tr}
\fmf{bg_fermion}{m,br}
\end{fmfgraph}
}}}
\vcenter{\vbox{\hbox{
\begin{fmfgraph}(80,50)
\fmfpen{.5pt}
\fmftop{ct}
\fmfbottom{cb}
\fmf{tight_fermion,left=.9}{ct,m,cb}
\fmf{tight_fermion,right=.9}{ct,m,cb}
\fmffreeze
\fmffixed{(0,0)}{ct,ct2}
\fmffixed{(0,0)}{cb,cb2}
\end{fmfgraph}
}}} \; &= \;\; \frac{1}{192}\widehat F_{mnpq} K^{mn} K^{pq} \Psi^{rstu}  
\widehat F_{rstu}\, ,\\
 & \\
\vcenter{\vbox{\hbox{
\begin{fmfgraph}(30,30)
\fmfpen{.5pt}
\fmfleft{tl,bl}
\fmfright{tr,br}
\fmf{bg_fermion}{m,tl}
\fmf{bg_fermion}{m,bl}
\fmf{bg_fermion}{m,tr}
\fmf{bg_fermion}{m,br}
\end{fmfgraph}
}}}
\vcenter{\vbox{\hbox{
\begin{fmfgraph}(80,50)
\fmfpen{.5pt}
\fmftop{ct}
\fmfbottom{cb}
\fmf{single_dotted_rev,right=.9}{ct,m}
\fmf{single_dotted,right=.9}{m,ct}
\fmf{tight_wiggly,right=.9}{cb,m,cb}
\fmffreeze
\fmffixed{(0,0)}{ct,ct2}
\fmffixed{(0,0)}{cb,cb2}
\end{fmfgraph}
}}} &= \;\; 4\, R\, \widehat F_{rstu} \Psi^{rstu}\, .
\end{aligned}
\end{equation*}
\caption{Disconnected two-loop diagrams contributing to the path integral at order
$\beta^0$. Note that the result for the diagram on the third line
includes graphs obtained by permutation of the time derivatives. One must also
add to this diagram the corresponding ghost diagrams which cancel the divergence
$\,\,\displaystyle \dot{}\!\Delta\!\dot{}\,(s,s)=\frac{{\rm d}^2}{{\rm
d}s {\rm d} t}\Delta(s,t)\Big\vert_{s=t}\,=\, \delta(0)-1$.}
\label{t:twoloop}
\end{table}

\end{fmffile}

\section{Boundary contributions to the index with torsion}
\label{s:boundary}
\subsection{The APS boundary correction with torsion}
\label{ss:travolta}

On manifolds with boundary, the central question is which boundary
conditions one must impose for spinors in order that the index problem
is well posed. Exactly this problem was solved by Atiyah, Patodi and
Singer in the mid-seventies~\cite{atiy1} for manifolds which have a
product form near the boundary ({\it i.e.}~when the manifold is a cylinder
with the boundary manifold as cross section). In this case the Hilbert space is
sufficiently under control and consistent boundary conditions can be
imposed on spinors.
Moreover, due to
the fact that the radial factor can be written down explicitly, the
index computation can be reduced to the problem of suitably counting the
difference between the number of positive and negative eigenvalues of
the boundary Dirac operator. 

For manifolds that are not of a product form, a split of the Hilbert
space is not as obvious. Nevertheless, due to the topological
nature of the index, it is possible to deform these manifolds until
their boundary takes a product form. The index then splits into a bulk
part plus the APS boundary term, amended by a modification due to the
deformation. This modification was pointed out by Gilkey~\cite{gilk1}
and we will generalise it here to also include torsion.

Let us now explain these matters in more detail. Suppose
$\partial=\partial/\partial r$ is a vector normal to the boundary
$\partial\cal M$ of the manifold in question. Then the Dirac operator
(possibly including a coupling to the antisymmetric part of the
contortion, assumed to have only components parallel to the boundary) takes the form
\begin{equation}
\label{e:Kurt}
\slashed{D}(e,A)=\gamma^{[r]}\Big[\left(\partial+\tfrac{1}{2}\omega(e)^\rho{}_{\rho[r]}\right)\,+\,
{\cal B}\, \Big]
\end{equation}
where ${\cal B}$ evaluated at the boundary $r=r_b$ is the boundary Dirac operator. We denote the curved
index in the normal direction spanned by the coordinate $r$ as $[r]$
and $\omega(e)^\rho{}_{\rho[r]}$ is the trace of the torsion-free part
of the spin connection computed from the vierbein $e_\mu{}^r$ (no
confusion should arise between the flat index $r$ and the radial
coordinate $r$).  Observe that
since $\slashed{D}(e,A)$ is antihermitean with respect to the volume
element $\sqrt{g(e)}$ the coefficient of the single outward pointing
Dirac matrix $\gamma^{[r]}=\gamma^r e_r{}^{[r]}$ is fixed to be that
appearing in~\eqn{e:Kurt}.  However, to verify that the Dirac operator
is antihermitean with respect to the inner product
$(\Xi,\Psi)=\int_{\cal M}\!{\rm d}^4x \sqrt{g(e)}\,\, \Xi^\dagger\,
\Psi$, along with taking into account differentiations by parts that
hit the measure factor $\sqrt{g(e)}$ one must also ensure that any
surface terms also vanish by imposing appropriate boundary conditions
on spinors. However, local Neumann or Dirichlet boundary conditions do
not lead to solutions of the first order homogeneous zero-mode Dirac
equation. Instead non-local boundary conditions are required.

Let us write the Dirac operator~\eqn{e:Kurt} in a 
chiral basis for the Dirac matrices
\begin{equation}
\label{e:chiral}
\slashed{D}=
\frac{1}{V}\begin{pmatrix}
  0  &  \left( \partial+\frac{1}{2}\omega(e)^\rho{}_{\rho[r]} \right) + B \\
  \left(\partial+\frac{1}{2}\omega(e)^\rho{}_{\rho[r]} \right)-B  &  0
  \end{pmatrix}
\end{equation}
where $V$ is some function of the normal variable $r$. The
\mbox{$2\times 2$} matrix $B$ is hermitean acting on spinors on the
compact boundary manifold $\partial\cal M$. The APS split of the index
computation into a bulk piece and a boundary piece can only be achieved
if~\eqn{e:chiral} takes a simpler form near the boundary. First of
all, $B$ has to be independent of $r$ and the connection trace $\omega(e)^\rho{}_{\rho[r]}$
should vanish. This holds when the manifold is a product metric
near the boundary. In addition we will require  that
there is no ``radial'' component of the torsion
({\it i.e.}~$A_{[r]\nu\rho}=0$) as this would enter the Dirac operator in a
way similar to the connection trace and prohibit an explicit solution
of the radial part of the eigenfunction. Although the APS condition
can be imposed by smoothly deforming the manifold near the boundary,
it is not obvious that such a deformation argument can be used to
set a radial torsion component to zero.

When these conditions are satisfied, Atiyah {\it et al.}~found that
one can consistently impose
the following non-local boundary conditions on spinors $\Psi$,
\begin{equation}
\label{e:Cobain}
\left.
\begin{aligned}
\sum_{l\geq 0} |l\,\rangle\langle\, l|\,\phi(r_b) &=0 \\[1em]
\sum_{l< 0}|l\,\rangle\langle \, l|\,\chi(r_b) &=0
\end{aligned}
\right\}\quad
\Psi=\left(
\begin{aligned}
\phi\\\chi
\end{aligned}\,
\right)\, ,
\end{equation}
where the boundary is at $r=r_b$ and
the two component spinors $| l\,\rangle$ are eigenspinors of $B$, 
\begin{equation}
B\, |l\,\rangle=l\,|l\, \rangle\, .
\end{equation}
Only for such non-local boundary conditions does a non-trivial set of
zero-modes of the Dirac operator survive. The index splits in a bulk
and a boundary piece as indicated according to
\begin{equation}
\label{e:BB}
{\rm index}\slashed{D}={\rm index(bulk)}
\,-\,
\frac{1}{2}\Big(\eta_B(0)+h_B\Big)\, 
\end{equation}
where ${\rm index(bulk)}$ is the usual integral over the second
Pontrjagin class, $h$ is the number of harmonic spinors on the
boundary manifold $\partial {\cal M}$, {\it i.e.}~the number of
solutions to $B\, \Psi = 0$, and $\eta_B(s)$ is a spectral invariant
of the boundary operator $B$ given by
\begin{equation}
\label{e:eta}
\eta_B(s)=\sum_{l\neq 0}\frac{{\rm sign}(l)}{|l|^s}
\end{equation}
where the sum runs over the eigenvalues $l$ of the boundary Dirac
operator $B$. As a function of the complex variable $s$, $\eta_B(s)$
has a well defined analytic continuation to $s=0$~\cite{atiy1,atiy2,atiy3}.

When $B$ only has a finite number of eigenvalues one can take $s=0$
before summing~\eqn{e:eta} (for other cases in which the APS
conditions can be made explicit see {\it e.g.}~\cite{hort1,nino1}). In this
case, one can understand the appearance of $\eta_B(0)$ in the expression
for the index rather easily. Assuming that near the boundary at $r=r_b$ 
the manifold is a product metric, the zero mode equation on this
cylinder can be written as
\begin{equation}
\begin{pmatrix}
0 & \partial + B \\[.5em]
\partial - B  & 0 
\end{pmatrix}
\begin{pmatrix}
 c_k\, e^{kr} | l \rangle \hfill \\[.5em]
 d_k\, e^{kr} | l \rangle 
\end{pmatrix}
=
\begin{pmatrix}
 (k+l)\, d_k\, e^{kr} | l \rangle \hfill \\[.5em]
 (k-l)\, c_k\, e^{kr} | l \rangle 
\end{pmatrix} = 0 \, 
\end{equation}
(for the product metric we may simultaneously diagonalise $B$ and
$\partial$).  This implies that positive chirality zero-modes (those
for which only the upper components are non-vanishing) have $k=l$
while those of negative chirality have $k=-l$. The boundary
conditions~\eqn{e:Cobain} imply that the positive chirality states
have only negative eigenvalues of $B$ while the negative chirality
states carry positive $B$-eigenvalues (due to the relation between $k$
and $l$ this implies that only zero-modes which vanish exponentially
as $r\rightarrow\infty$ remain). The contribution to the index from
the cylinder is therefore the disparity between the number of negative
and positive eigenvalues of $B$ which is clearly given by~\eqn{e:eta}
at $s=0$ and this limit may be taken before performing the finite
summation (the factor $1/2$ in~\eqn{e:BB} arises from the fact that
half of the spinor boundary modes do not continue smoothly into the
interior~\cite{atiy1,hort1}). The APS theorem renders this argument
rigorous for operators $B$ with an infinite number of eigenvalues and
also shows that the index problem indeed splits in a bulk and boundary
piece according to~\eqn{e:BB} by appropriately gluing the two pieces
of the manifold.

Before we discuss the Gilkey correction term, we will now first show
how the $\eta$ invariant can be computed for a class of boundary Dirac
operators general enough to include the torsion-full examples to be
discussed later.

\subsection{Computation of the $\eta$ invariant for manifolds with
torsion and $S^3$ boundary}
\label{ss:eta_s3}

The $\eta$ invariant must be computed on a case by case basis and
since it involves all non-zero boundary eigenvalues this can be rather
tricky. Fortunately this computation has been carried out by Hitchin
for a large class of boundary geometries~\cite{hitc2} (to be specific,
squashed $S^3$ metrics). The addition of torsion modifies the boundary
Dirac operator by what amounts to only a slight generalisation of
Hitchin's calculations.  However, in keeping with the rest of this
paper and because the explicit solutions we find to the boundary
Dirac equation will be needed when we explain how one is able to
compare index problems for manifolds with mutually anticommuting Dirac
operators in section~\ref{s:dual_operators}, we will now present these
additional details.

We begin with a boundary operator of the form
\begin{equation}
\label{e:macartney}
{\cal B} =
  S\left(\gamma^{34} \Sigma_{1} + \gamma^{42} \Sigma_{2}\right)
 + \frac{1}{\lambda} \gamma^{14} \Sigma_{3} 
 - \gamma^{1234}\left[ \frac{\lambda^2+2}{4\lambda}+Z\right] \, .
\end{equation}
where $\lambda$, $S$ and $Z$ are arbitrary parameters
(depending, in general, on $r_b$). The operators
$\Sigma_i$ commute with the $SO(3)$ Killing vectors ({\it i.e.} they are the
inverses of the left-invariant forms) and satisfy the algebra
\begin{equation}
[\Sigma_1,\Sigma_2]=\Sigma_3 \quad \text{and cyclic.}
\end{equation}
To motivate the somewhat obscure looking form we have chosen for the
operator~\eqn{e:macartney}, we note that it is exactly the case solved by
Hitchin~\cite{hitc2} when $S=Z=0$. The generalised form presented here
can be
handled too and (as we will see) appears in the example manifolds 
discussed in section~\ref{s:dual_operators}. 
A useful basis for the Dirac matrices is
\begin{equation}
\label{e:what's in a basis?}
\gamma^1=\left(\begin{array}{cc}0\,\,&\,\,{\mathbb 1}\\ {\mathbb
1}\,\,&\,\,0
\end{array}\right)
\quad
\gamma^2=\left(\begin{array}{cc}0&-i\sigma_x\\ i\sigma_x&0\end{array}\right)\quad
\gamma^3=\left(\begin{array}{cc}0&-i\sigma_y\\ i\sigma_y&0\end{array}\right)\quad
\gamma^4=\left(\begin{array}{cc}0&-i\sigma_z\\ i\sigma_z&0\end{array}\right)\quad
\end{equation}
One then has
\begin{equation}
{\cal B}=\begin{pmatrix} P(S,Z) & 0        \\ 
                         0      & -P(-S,Z) 
         \end{pmatrix}\, ,
\end{equation}
where the ``Hitchin--operator'' is given by 
\begin{equation}
P(S,Z)=iS[\Sigma_1\sigma_x+\Sigma_2\sigma_y]+\frac{i}{\lambda}\Sigma_3\sigma_z
+\frac{\lambda^2+2}{4\lambda}+Z\, .
\end{equation}
Therefore it is most useful to solve the eigenvalue problem
$P(S,Z)\,\varphi=z\varphi$. To begin with call $\Sigma_i=-iJ_i$ where
$J_i$ is the angular momentum operator $(i=1,..,3)$ (and we
employ the usual eigenstates $|j,m\rangle$ of $J^2$ and $J_3$).  Note
that the Hitchin operator commutes with the Casimir $[J^2,P(S,Z)]=0$
so we may look for solutions at a fixed $j$. Actually, it is easy to
solve this equation with the following ansatz
\begin{equation}
\varphi=\left(\begin{array}{c} \alpha\, |j,m\rangle \\{} |j,m+1\rangle 
\end{array}\right)
\end{equation}
where we define $|j,j+1\rangle=0=|j,-j-1\rangle$. Then if $m=j$ or $m=-j-1$
 we clearly obtain solutions with eigenvalue
\begin{equation} z=j/\lambda +(\lambda^2+2)/4\lambda\,+\,Z.
\end{equation}
Now if $j>m>-j-1$ so that $\Delta\equiv\sqrt{(j-m)(j+m+1)}$ is real and non-vanishing
then we have solutions with eigenvalues
\begin{equation}
z_\pm=\alpha_\pm\,S\Delta-(m+1)/\lambda
+(\lambda^2+2)/4{\lambda}+Z
\end{equation}
where $\alpha_\pm$ solves the equation
\begin{equation}
\alpha_\pm^2-\frac{2m+1}{\Delta S\lambda}\,\alpha_\pm-1=0\, ,
\end{equation}
so that explicitly
\begin{align}
z_\pm&=
\frac{\lambda^2\pm2\sqrt{(2m+1)^2 + 4 S^2\lambda^2(j-m)(j+m+1)}}{4\lambda}
\,+\,Z\, .
\end{align}
To conform with the notation of Hitchin we relabel $j=(p+q-1)/2$
and $m=(p-q-1)/2$ so that the eigenvectors, eigenvalues and their multiplicities
read
\begin{equation}
\label{e:spectrum}
\begin{aligned}
~ &\text{\underline{\underline{eigenvector}}} &&\text{\underline{\underline{eigenvalue}}} &&\text{\underline{\underline{multiplicity}}}\\[1em]
~ & \left(
  \begin{array}{c}
  |\frac{p-1}{2},\frac{p-1}{2}\rangle\\0
  \end{array}
  \right)\, , \quad
  \left(
  \begin{array}{c}
  0\\|\frac{p-1}{2},\frac{1-p}{2}\rangle
  \end{array}
  \right)\quad
&& \frac{p}{2\lambda}+
  \frac{\lambda}{4}+Z \quad
&& 2p \\[1em]
~ & \left(
  \begin{array}{c}
  \frac{p-q\pm\sqrt{4pqS^2\lambda^2+(p-q)^2}}{2\lambda\sqrt{pq}}
  |\frac{p+q-1}{2},\frac{p-q-1}{2}\rangle\\|\frac{p+q-1}{2},\frac{p-q+1}{2}\rangle
  \end{array}
  \right)\quad\quad\quad\!\!\!
&& \frac{\lambda}{4}\pm\frac{\sqrt{4pqS^2\lambda^2+(p-q)^2}}{2\lambda}+Z
  \quad\quad\quad\!\!\!
&& p+q
\end{aligned}
\end{equation}
where the integers $p$ and $q$ take all values strictly greater than zero
(note that the multiplicities above refer to distinct eigenvalues). 
To understand the multiplicities quoted in~\eqn{e:spectrum}, it is
useful to briefly explain the representation theory of states on
$S^3$. The metric on the squashed $S^3$ (considered in this section)
may be represented (up to an overall irrelevant scaling) in terms of
left invariant forms $\Sigma^i$ ($i=1,..,3$) as
\begin{equation}
{\rm d}s^2\big\vert_{S^3} = (\Sigma^1)^2+(\Sigma^2)^2+(\Sigma^3/(S\lambda))^2
\end{equation}
{}From the invariant forms one may write down left invariant vectors 
$\Sigma_i$ which commute with the Killing vectors $k_i$
and satisfy the algebra (the explicit forms of these vectors
are given in section~\ref{ss:taub-nut})
\begin{equation}
\begin{aligned}
{}[\Sigma_1,\Sigma_2]  &= \Sigma_3 \quad & \text{and cyclic,} \\
[\Sigma_i\,,k_j]       &= 0 \, , &\\
[\,\!k_1\,,k_2]            &= k_3      \quad & \text{and cyclic.}
\end{aligned}
\end{equation}
Representing the operators $\Sigma_i$ on the space $\{|j,m\rangle\}$
as above and the Killing vectors on a second copy of this space
$\{|J,M\rangle\}$ then from the relation
\begin{equation}
\sum_{i=1}^3 (\Sigma_i)^2 = \sum_{i=1}^3 (k_i)^2
\end{equation} 
expressing equality of the Casimirs on $S^3$ (which may be easily
verified
from the explicit forms in~\ref{ss:taub-nut}), one finds
the representation space of functions on $S^3$ to be
$\{|j,m\rangle\otimes|j,M\rangle\}$ which yields the multiplicities quoted above
(since in~\eqn{e:spectrum} we suppressed the dependence on the 
``right representation'' $|J,M\rangle$).

We are now left with the task of computing $\eta_B(0)=\eta_P(0)$ as
in~\eqn{e:eta} given the spectrum~\eqn{e:spectrum}.  Therefore we must
find the analytic continuation to $s=0$ of (we drop an overall factor
$(2\lambda)^{s}$ which clearly does not modify $\eta_B(0)$)
\begin{equation}
\begin{aligned}
\eta_B(s) &= \sum_{p>0} \frac{2p}{(p+\lambda^2/2+2Z\lambda)^{^{\textstyle s}}} \\
        &\quad+ \sum_{p,q>0}\,\left\{ \frac{p+q}{\left( \lambda^2/2 + 
          \sqrt{D(\lambda S,p,q)}+2Z\lambda\right)^{\textstyle s}} - 
\frac{p+q}{\left(-\lambda^2/2 + \sqrt{D(\lambda S,p,q)}-2Z\lambda\right)^{\textstyle s}}
\,\right\}
\end{aligned}
\end{equation}
where $D(\lambda S,p,q)=4pq(\lambda S)^2 + (p-q)^2$. (Note that in all
cases that we are interested, $\lambda$, $S$ and $Z$ are such that
$-\lambda^2/2 + \sqrt{D(\lambda S,p,q)}-2Z\lambda>0$, which motivates
the above split in a sum over positive and negative eigenvalues, but
in general one should of course worry about this separation as well as
any zero-modes of the boundary operator).  The single sum can be
expressed in zeta functions without any difficulty. The double sum
requires a lengthy analysis, most easily performed by first Taylor
expanding in powers of the new variable $\Lambda$ defined by
$\Lambda^2/2=\lambda^2/2+2Z\lambda$,
\begin{equation}
\begin{aligned}
\eta_B(s) &= 2\,\zeta(s-1,\Lambda^2/2) - \Lambda^2\, \zeta(s,\Lambda^2/2) \\
        &\quad -2s\left(\frac{\Lambda^2}{2}\right) f\big((s+1)/2\big) -
        \frac{2s(s+1)(s+2)}{3!} \left(\frac{\Lambda^2}{2}\right)^3 
		f\big((s+3)/2\big) + \cdots
\end{aligned}
\end{equation}
where the dots vanish for $s=0$ and the function $f(s)$ is defined by
\begin{equation}
f(s) = \sum_{p,q>0} \frac{p+q}{D(\lambda S,p,q)^{\textstyle s}}\, .
\end{equation}
The terms involving the
zeta functions yield $\Lambda^4/4 - 1/6$ while the
analysis of the residues of $f(s)$ is rather nontrivial and produces~\cite{hitc3}
\begin{equation}
\begin{aligned}
{\rm res}\left[ f\left(\frac{s+1}{2}\right) \right]_{s=0} &=  \frac{1}{3}\left((\lambda S)^2 -1 \right) \, ,\\
{\rm res}\left[ f\left(\frac{s+3}{2}\right) \right]_{s=0} &=  \frac{1}{(\lambda S)^2} \, .
\end{aligned}
\end{equation}
This finally yields the $\eta$ invariant at $s=0$
\begin{equation}
\label{e:eta_general}
\begin{aligned}
\eta_B(0)&=-\frac{1}{6} 
        -\frac{\lambda}{3}\bigg[ \frac{16\, Z^3}{S^2} - 4\, Z \bigg]
        +\frac{\lambda^2}{3} \bigg[ 12\,Z^2 - \frac{12\, Z^2}{S^2} +
        1\, \bigg] \\[1em]
 &\quad -\frac{\lambda^3}{3}\bigg[ \frac{3\, Z}{S^2} + 4\, S^2 Z  -
        6\, Z \bigg] 
        +\frac{\lambda^4}{4}\bigg[ 1 - \frac{4\, S^2}{3} -
        \frac{1}{3\,S^2} \bigg] \, .
\end{aligned}
\end{equation}
At $S=1$ and $Z=0$, we recover Hitchin's result
$\eta_B(0)=-1/6+\lambda^2/3-\lambda^4/6$.  So summarising, our final
result for the $\eta$ invariant for the class of Dirac operators
in~\eqn{e:macartney} is given by the above result~\eqn{e:eta_general}.

\subsection{Boundary correction for non-product metrics}
\label{e:gilkeybound}

The APS analysis presented in the previous two subsections is valid
for the manifolds whose metric takes a product form~\cite{atiy1} at
the boundary (more intuitively, those manifolds which, near the
boundary, are a cylinder with the boundary manifold as cross
section). If this is not the case, the manifold has to be deformed to
such a product structure near the boundary before the above machinery can be
applied. The bulk term should in turn also be computed for this {\em
deformed}\/ metric. As an explicit form of the deformed metric is
often not available, it is, however, much easier to determine instead
the error that one has made by computing the bulk contribution using
the original metric. This error term, due to Gilkey~\cite{gilk1}
for torsion-less manifolds, has to be subtracted from the bulk.

The form of this correction is rather simple. In the absence of
torsion, the bulk term can be written as a boundary integral
\begin{equation}
\frac{1}{24.8\pi^2}\,\int_{\cal M} R\wedge R = \frac{1}{24.8\pi^2}\,\int_{{\cal M}} {\rm d}C = \frac{1}{24.8\pi^2}\,\int_{\partial 
  {\cal M}} C\, ,
\end{equation}
provided discontinuities of $C$ are taken into account properly. For
the metric deformed to a product structure near the boundary, we get a
boundary integral with $C$ replaced by the Chern--Simons term of the
product metric $g_{\mu\nu}^\sharp=g_{\mu\nu}(r=r_b)$, 
$C^\sharp$. Therefore, in the absence of torsion, the
correction term that has to be added is
\begin{equation}
\label{e:sinatra}
\frac{1}{24.8 \pi^2}\int_{\partial {\cal M}}[C^\sharp -C]\, ,
\end{equation}
where the Chern--Simons three-form $C=\omega\wedge
R-\frac{1}{3}\omega\wedge\omega\wedge\omega$ (and similarly for
$C^\sharp$ built from ``sharped'' objects computed from the product
metric). Note that the integrand of~\eqn{e:sinatra} can be written as
$\theta\wedge R$ where $\theta=\omega-\omega^\sharp $ is the second
fundamental form and $\omega$ and $\omega^\sharp $ are the
torsion-free spin
connections computed from the metric and product metrics respectively.
(Where we {\em warn}\/ the reader that from here on we replace $\omega(e)$ by
simply $\omega$ to denote the torsion-free spin connection.)
However, for intuitive, along with practical reasons, we prefer the
form given in~\eqn{e:sinatra}.

For the torsion--full case the integrand of the bulk index derived in
section~\ref{s:bulk} above may be written as the exterior derivative 
of a generalised Chern--Simons form
\begin{equation}
\label{e:franks_dog_was_diamond}
\omega\wedge R-\frac{1}{3}\omega\wedge\omega\wedge\omega -\frac{1}{2}A\wedge F-2\, {\cal K}
\,\equiv\,
C(A)-2\, {\cal K}
\end{equation}
where the three-form ${\cal K}$ is the Hodge dual of the vector ${\cal
K}^\mu$ in~\eqn{e:no_isaacs_dog_was_diamond},
\begin{equation}
\label{e:twelvebar}
{\cal K}^\mu = \frac{1}{\sqrt{g}}\epsilon^{\mu\nu\rho\sigma} {\cal
K}_{\nu\rho\sigma}\, .
\end{equation}
Therefore, the boundary
correction for the torsion-full case is simply 
\begin{equation}
\label{e:do_we_really_need_sinatra_twice?}
\frac{1}{24.8 \pi^2}\int_{\partial {\cal M}}[C^\sharp(A)  -2\,{\cal
K}^\sharp -C(A)+2\,{\cal K}]\, ,
\end{equation}
where ${\cal K}^\mu$ is computed by inserting $r=r_b$ in the axial
contortion vector $A^\mu$
before computing the covariant derivatives in~\eqn{e:no_isaacs_dog_was_diamond}.

\subsection{Generalised APS index theorem}
\label{ss:all_together}

Orchestrating the above  bulk results presented
in~\eqn{e:bulkresult} as well as the $\eta$ term and the boundary
contribution~\eqn{e:do_we_really_need_sinatra_twice?} we obtain an
APS index theorem generalised to manifolds with torsion
\begin{equation}
\label{e:johnny} 
\begin{aligned}
{\rm index}\slashed{D}(e,A)&=
\frac{1}{24.8\pi^2}\int_{{\cal M}}\,\Big[
 R( e)^{mn}\wedge R( e)_{nm}
- \frac{1}{2} F(A)\wedge F(A) -2\, \sqrt{g}\, D_\mu {\cal K}^\mu
\Big] \\[1em]
&\quad +\,
\frac{1}{24.8\pi^2}\int_{\partial{\cal M}} \Big[ C^\sharp(A)
-2\,{\cal K}^\sharp-
C(A)+2\,{\cal K} \Big] \\[1em]
&\quad \,-\,
\frac{1}{2}\big[\eta_B(0)
+h_B\big] \, ,
\end{aligned}
\end{equation}
with the three-form ${\cal K}$ defined by~\eqn{e:twelvebar}
and~\eqn{e:no_isaacs_dog_was_diamond} above. 
A result similar to~\eqn{e:johnny} has appeared in a string theoretical
context in~\cite{bell1}
for which one encounters
exact torsion and most of the terms above vanish. 
Needless to say, in the general case presented here there are many 
extra subtleties as discussed in the preceeding text.

\section{Index theorems for manifolds with mutually anticommuting 
Dirac operators}
\label{s:dual_operators}
\subsection{General formulation}
\label{ss:hermiticity}

The existence of mutually anticommuting Dirac operators on a pair of
manifolds can be employed to derive a relation between the index
theorems on these manifolds which we shall now present in detail.  At
the end of this section we test this relation on the explicit example
of the Taub--NUT and dual Taub--NUT manifolds.

To begin with, given a manifold ${\cal M}$ with (possibly torsion-full)
Dirac 
operator $\slashed{D}(e,A)$ satisfying
\begin{equation}
\slashed{D}(e,A)=-\slashed{D}(e,A)^\dagger\, , \quad 
\{\slashed{D}(e,A),\gamma^5\}\, = \, 0\, ,
\end{equation}
then 
let us assume that we have found a second operator $\slashed{D}(\tilde e,\widetilde V)$
acting on spinors defined on ${\cal M}$
which solves the equations
\begin{gather}
\label{e:cat}
\{\slashed{D}(e,A),\slashed{D}(\tilde e,\widetilde V)\} = 0 \\[.5em]
\label{e:Jimi}
\slashed{D}(\tilde e,\widetilde V)=-\slashed{D}(\tilde e,\widetilde V)^\dagger\, , 
\quad \{\slashed{D}(\tilde e,\widetilde V),\gamma^5\}\, = \, 0\, .
\end{gather}
At this point we do {\em not}\/ yet identify $\slashed{D}(\tilde
e,\widetilde V)$ with the Dirac operator on a manifold $\widetilde{\cal
M}$. Importantly the hermiticity requirement in~\eqn{e:Jimi} is
formulated on the manifold $\cal M$ (the precise definition of the
adjoint operation for spinors on manifolds with boundary was presented
in section~\ref{ss:travolta}).  In general, given that the inverse
vierbeine $e_r{}^\mu\neq \tilde e_r{}^\mu$, observe that it will
certainly be necessary to include the additional vector coupling
$\widetilde V^\mu$ in $\slashed{D}(\tilde e,\widetilde V)$ to satisfy
both~\eqn{e:cat} and~\eqn{e:Jimi}.

As the operators $-i\slashed{D}(\tilde e,\widetilde V)$ and $-i\slashed{D}(e,A)$ 
anticommute, we can construct
a new operator $\gamma^5 \slashed{D}(e,A)$ 
commuting with the
original Dirac operator,
\begin{equation}
\label{e:alg}
{}[ -i\slashed{D}(\tilde e,\widetilde V), \gamma^5\slashed{D}(e,A)] = 0\, .
\end{equation}
These two operators (hermitean on ${\cal M}$) can thus be diagonalised simultaneously.
Eigenspinors with non-zero eigenvalues occur in pairs for
both operators, since they anticommute with $\gamma^5$.
However, eigenspinors with vanishing
$-i\slashed{D}(\tilde e,\widetilde V)$ eigenvalues
are not 
necessarily those with zero $\gamma^5\slashed{D}(e,A)$ eigenvalues.
Nonetheless, in the
computation of the index,
\begin{equation}
{\rm index}\,\slashed{D}(\tilde e,\widetilde V) = n_+(\slashed{D}(\tilde e,\widetilde V)) - n_-(\slashed{D}(\tilde e,\widetilde V))\, ,
\end{equation}
they still occur in pairs ($n_+$ denotes the number of zero-modes with 
positive chirality). The above expression therefore
only receives contributions from those zero-modes that are also
zero-modes of $\gamma^5\slashed{D}(e,A)$, that is 
${\rm Ker}{\hspace{.5mm}}^\prime \,  (-i\slashed{D}(\tilde e,\widetilde V)) \subset {\rm
Ker}{\hspace{.5mm}}^\prime \,  \, (\gamma^5\slashed{D}(e,A))$, where
${\rm Ker}{\hspace{.5mm}}^\prime \,  $ denotes the set of zero-modes 
which do not have a partner of opposite chirality. 
The same holds true when we interchange the r\^ole of the two
Dirac operators and we therefore also have 
${\rm Ker}{\hspace{.5mm}}^\prime \,  (-i\slashed{D}(\tilde e,\widetilde V)) \supset {\rm Ker}{\hspace{.5mm}}^\prime \,  (\gamma^5\slashed{D}(e,A))$. 
We may therefore conclude that
the indices of $-i\slashed{D}(\tilde e,\widetilde V)$ and $\gamma^5 \slashed{D}(e,A)$ are equal 
and hence\footnote{The above argument was pointed out to us by
Jan-Willem van Holten~\cite{holt5}.}
\begin{equation}
\label{e:Hendrix}
{\rm index}\,\slashed{D}(\tilde e,\widetilde V)={\rm index}\, \slashed{D}(e,A)\, ,
\end{equation}
since if $\Psi$ is an eigenspinor of the operator
$\gamma^5\,\slashed{D}(e,A)$ then $\gamma^5 \Psi$ is the corresponding
eigenspinor of $-i\slashed{D}(e,A)$.

The equation~\eqn{e:Hendrix} becomes much more interesting if one is
able to view the operator $\slashed{D}(\tilde e,\widetilde V)$ as the
Dirac operator on the manifold $\widetilde{\cal M}$ but one must then
carefully consider the space on which the operators 
$\slashed{D}(e,A)$ and $\slashed{D}(\tilde e,\widetilde V)$ act.
The reason being that typically
one is interesting in computing the index of the Dirac operator as a trace
over states
\begin{equation}
{\rm index} \, \slashed{D}\, = \, 
{\rm Tr}\left[ \gamma^5\,\exp
\Big(\frac{\textstyle\beta}{\textstyle2}\slashed{D}^2\Big)\right]
\end{equation}
where the trace extends over the space of spinors living on the manifold 
of interest. However, we do not want to compute the index of the
second Dirac operator $\slashed{D}(\tilde e,\widetilde V)$ on the manifold $\cal M$,
but rather on the new manifold $\widetilde{\cal M}$ obtained by expressing
$\slashed{D}(\tilde e,\widetilde V)=\gamma^r\,\tilde e_r{}^\mu\, \partial_\mu\,+\cdots$ and 
regarding $\tilde e_r{}^\mu$ as the inverse vierbein. Only in this way
do we obtain a relation between index theorems on independent 
manifolds $\cal M$
and $\widetilde{\cal M}$.
{\em A priori\/}, however, there is no reason to expect that the space of spinors
on $\cal M$
should coincide with that on $\widetilde{\cal M}$.
Furthermore, although the operator 
$\slashed{D}(\tilde e,\widetilde V)$ was assumed to be antihermitean on the manifold
$\cal M$, since the volume elements of the two manifolds will in general
not be equal, the operator $\slashed{D}(\tilde e,\widetilde V)$ will have no definite
hermiticity viewed as an operator on spinors on $\widetilde{\cal M}$.
Therefore we make the additional assumption, (which holds for the
example we 
have in mind)

\vspace{.3cm}
\noindent{\bf Assumption:} 
The space of spinors (which were defined in detail in subsection~\ref{ss:travolta} 
in such a way that
index problems are well-posed) viewed as a four-component space of functions
on $\cal M$ and $\widetilde{\cal M}$ coincide.

\vspace{.3cm}
\noindent
Notice that as a consequence the index problem for the operator
$\slashed{D}(\tilde e,\widetilde V)$ is well posed on the manifold
$\widetilde{\cal M}$. This can be seen as follows. Even though the
operator $\slashed{D}(\tilde e,\widetilde V)$ has no definite hermiticity on the
manifold $\widetilde{\cal M}$ the operator
$\widetilde{\slashed{D}}(\tilde e,\widetilde V)\equiv\sqrt{g(e)/g(\tilde
e)}\,\slashed{D}(\tilde e,\widetilde V)\,\sqrt{g(\tilde e)/g(e)}$ is anti-hermitean
on $\widetilde{\cal M}$. Furthermore if $\Psi_\lambda$ are a complete
orthonormal set of eigenfunctions of $\slashed{D}(\tilde e,\widetilde V)$ on
${\cal M}$ then $\sqrt{g(e)/g(\tilde e)}\Psi_\lambda$ are a complete
set of orthonormal eigenfunctions of $\widetilde{\slashed{D}}(\tilde
e,\widetilde V)$ on $\widetilde{\cal M}$. Clearly however, by the 
assumption, if the above similarity transformation is non-singular
then the set of functions $\Psi_\lambda$ are still complete (although
no longer orthonormal) on $\widetilde{\cal M}$ so that
$\slashed{D}(\tilde e,\widetilde V)$ can still be diagonalised on
$\widetilde{\cal M}$ leading to a well posed index problem on that manifold.
 
It should now be clear that given the above assumption
we have equality of well posed index problems
calculated 
on $\cal M$ and 
$\widetilde{\cal M}$,
\begin{equation}
\label{e:King}
{\rm index}\, \slashed{D}(e,A)
\,\equiv\,
{\rm Tr}_{{}_{\!{}_{{\cal M}}}}\, 
\gamma^5 \, \exp \big(\frac{\beta}{2}\slashed{D}(e,A)^2\big)
\,=\,
{\rm Tr}_{_{\!_{\widetilde{\cal M}}}}\, 
\gamma^5 \, \exp \big(\frac{\beta}{2}\slashed{D}(\tilde e,\widetilde V)^2\big)
\,\equiv\,
{\rm index}\, \slashed{D}(\tilde e,\widetilde V)
\end{equation}
which is the main result of this section.
If we recall the discussion of boundary conditions for spinors on
manifolds with boundary in section~\ref{ss:travolta}, we see that to verify
our assumption for the manifolds ${\cal M}$ and
$\widetilde{\cal M}$ one only needs to require equality of the
projection operators
\begin{equation}
\label{e:proj}
\sum_{l\geq 0} |l\,\rangle\langle\, l|
\,=\,
\sum_{\, \tilde l  \,\geq 0} |\, \tilde l  \,\,\rangle\langle\, \, \tilde l  \,|
\end{equation} 
where $|l\,\rangle$ and $|\, \tilde l  \,\,\rangle$ are the 
eigenvalues of the respective boundary Dirac operators $B$ and
$\widetilde B$.
Remarkably, we find that~\eqn{e:proj} holds 
for the Taub--NUT and dual Taub--NUT  manifolds.  
   
Note that to compute the index of $\slashed{D}(\tilde e,\widetilde V)$ on
$\widetilde{\cal M}$ we can split the operator 
into a sum of antihermitean and hermitean
operators
on $\widetilde{\cal M}$ by 
writing
\begin{equation}
\label{e:Kiss}
\slashed{D}(\tilde e,\widetilde V)
\,\equiv\,{\slashed{D}}(\tilde e,0)\, +\, \gamma^\mu \widetilde V_\mu
\end{equation}
with
\begin{align}
\label{e:metallica}
\slashed{D}(\tilde e,0)&=
\Big[\slashed{D}(\tilde e,\widetilde V)-\frac{1}{2}\gamma^\mu
(\omega^\rho{}_{\rho\mu}-\omega(\tilde e)^\rho{}_{\rho\mu})\Big]\\
\widetilde V_\mu&=\frac{1}{2}\,\omega^\rho{}_{\rho\mu}
-\frac{1}{2}\,\omega(\tilde e)^\rho{}_{\rho\mu}
\end{align}
where $\omega(\tilde e)_{\mu rs}$ is the torsion free part of the spin
connection computed from the inverse vierbein $\tilde e_r{}^\mu$.  The trick
now is that the hermitean term $\gamma^\mu \widetilde V_\mu$ is nothing but the
coupling of the Dirac operator to a purely imaginary abelian gauge
field whose contribution to the index we may obtain by a naive
analytic continuation from the well known results for the index of the
Dirac operator with such couplings\footnote{In the example below we
will anyway find a vanishing contribution 
from this
additional abelian vector coupling, however in general it is quite
simple to handle. The bulk term is just the usual result for the
abelian anomaly modulo an
overall minus sign due to the analytic continuation. The Chern--Simons
and APS boundary corrections are computed exactly as described in
section~\ref{s:boundary}.}.

In the remainder of this section we show that manifolds with mutually
anticommuting Dirac operators can be found if invertible Killing--Yano
tensor exist and the above relation is realised explicitly by  
Taub--NUT and its dual manifold. Also provided are all relevant
geometric data for these manifolds.

\subsection{Killing--Yano dual manifolds}
\label{s:ky_duality}

Manifolds with torsion admitting two ``Dirac operators'' that satisfy
the properties discussed in the previous section do in fact exist and
are understood in a systematic way \cite{riet2} (at least in the
case where the Dirac operator is viewed as a classical supercharge). 
Before we move on to
an explicit index computation 
we describe
the Killing--Yano 
technique for  generating these manifolds.

The operator $\slashed{D}(\tilde e,\widetilde V)$ can be viewed as the
quantum analogue of the supercharge of a spinning particle and in this
context the additional Dirac operator $\slashed{D}(e,A)$ generates the
extended supersymmetry. Up to ordering ambiguities this means that the
conditions on the tensors $e_\mu{}^r$ and $A_{rst}$ appearing in the
additional Dirac operator such that it anticommutes with the original
Dirac operator can be deduced from the existence conditions of
extended supersymmetry of the classical model \cite{gibb1} (see
\cite{riet2} for the extension to include torsion). In short,
additional supercharges on the manifold $\widetilde{\cal M}$ are given by
\begin{equation}
\slashed{D}(e,A) = \gamma^r e_r{}^\mu  \Big( \partial_\mu + \tfrac{1}{4}\big(
\omega(\tilde e)_{\mu st} -
            \tfrac{1}{3} (e^{-1})_\mu{}^p c_{pst} \big)\gamma^{st}  
\Big)\, ,
\end{equation}
where the tensor $e_r{}^\mu$ satisfies
\begin{equation}
D(\tilde e)_{\mu} e_{\nu}{}^r + D(\tilde e)_{\nu} e_{\mu}{}^r = 0\, ,
\end{equation}
which is the Killing--Yano \cite{yano1} equation for $e_\nu{}^r$ on
the manifold $\widetilde{\cal M}$.
Here we have raised and lowered indices with the vierbein $\tilde
e_\mu{}^r$ on $\widetilde{\cal{M}}$ so that $e_\nu{}^r= \tilde
e_\nu{}^s e_s{}^\mu \tilde e_\mu{}^r$ and $D(\tilde e)_\mu$ denotes the
torsion-free covariant derivative on $\widetilde{\cal M}$. 
The connection is determined by	requiring
\begin{equation}
c_{rst} = D(\tilde e)_{[r} e_{st]}\, ,
\end{equation}
where again we have employed the vierbeine on $\widetilde{\cal M}$ to
flatten indices. Note however that when we come to regard $e_r{}^\mu$
as the inverse vierbein on the dual manifold $\cal{M}$ we will denote
its inverse ({\it i.e.}~the vierbein itself on $\cal{M}$) as
$e_\mu{}^r\equiv(e^{-1})_\mu{}^r$. Viewing $\omega(\tilde
e)_{\mu st}-\frac{1}{3}(e^{-1})_\mu{}^p c_{pst}$ as the connection on the dual
manifold $\cal{M}$ the  contortion on that manifold is then
\begin{equation}
K_{t \mu s}=\omega(\tilde e)_{\mu st}-\omega(e)_{\mu
st}-\frac{1}{3}(e^{-1})_\mu{}^p c_{pst}\, .
\end{equation}
the totally antisymmetric part (on $\cal M$) of which defines the
tensor $A_{rst}=K_{[rst]}$.

Indeed just as complex structures (with one flat index) can be used to define
a local Lorentz frame different from the frame spanned by the inverse
vierbein, an interpretation of the Killing--Yano tensor $e_r{}^\mu$ as
an inverse
vierbein is possible as well. In this case, however, the manifold for
which the inverse vierbein is $e_\mu{}^r$ is not identical to the
original one given by $\tilde e_\mu{}^r$ (the square $e_\mu{}^r e_{\nu
r}$ is a Killing tensor which does not normally coincide with the
metric on ${\cal M}$). This observation was first elucidated in
\cite{riet2} and plays a central r\^ole in this paper by providing an
example of manifolds with mutually anticommuting Dirac operators.

At the quantum level one must study the possible orderings when making
the transition from a classical supercharge to the Dirac operator
acting on a spinorial Hilbert space. For example, we have already  
discussed this problem in detail in section~\ref{s:bulk} (there  
symmetry principles such as general coordinate and local
Lorentz invariance of the index are central considerations). The key
observation of the previous subsection in this respect is the r\^ole
of the coupling to the trace of the torsion $\widetilde V^\mu$. Needless to say
exactly this coupling is absent in the classical Poisson bracket
formulation of the Killing--Yano technique.  Upon inclusion of the
torsion trace one must reanalyse the quantum anticommutator of Dirac
operators. In explicit examples we have found that in order to obtain
a vanishing anticommutator one must ensure that both
$\slashed{D}(\tilde e,\widetilde  V)$ and $\slashed{D}(e,A)$ are
antihermitean on the manifold in question which necessitates the addition
of the abovementioned torsion trace $\widetilde V^\mu$.

Unfortunately, manifolds which admit Killing--Yano tensors are not
nearly as well understood as their cousins that play a role in $D\geq
2$ extended supersymmetry: K\"ahler and hyper-K\"ahler manifolds. The
most extensive systematic study so far was made in
\cite{diet1,diet2}. Though very explicit, their analysis only concerns
the local geometrical structure of manifolds admitting Killing--Yano
tensors. Lacking a systematic topological analysis, this information
has to be extracted for every example being studied. For this reason,
we will in this paper focus on the Taub--NUT manifold
\cite{taub1,newm1} for all sample calculations. It should however be
noted that at least the Kerr metric is tractable as well; an extension
of those results to include the Kerr-Newman family could be used to
study a generalisation of our results to include electromagnetic
coupling, although there are no conceptual problems expected
there. The much simpler Schwarzschild metric only admits a {\em
non-invertible\/} Killing--Yano tensor, which therefore does not lead
to a non-singular inverse vierbein.

\subsection{Example: Taub--NUT geometry and its dual}
\label{ss:taub-nut}

Both Taub--NUT and dual Taub--NUT possess an $SO(3)$ isometry
generated by 
\begin{equation}
\label{e:killing_vectors}
\begin{aligned}
k_{1}{}^\mu \partial_\mu = k_{1}
   &= -\,\sin\phi\partial_{[\theta]} - \cos\phi\cot\theta \partial_{[\phi]}
      + \frac{\cos\phi}{\sin\theta}\partial_{[\psi]}\, ,\\
k_2 &= \,\cos\phi\partial_{[\theta]} - \sin\phi\cot\theta\partial_{[\phi]}
      + \frac{\sin\phi}{\sin\theta}\partial_{[\psi]}\, ,
\end{aligned}
\end{equation}
as well as the trivial vector
\begin{equation}
k_3 = \partial_{[\phi]}\, .
\end{equation}
(there is also a fourth trivial Killing vector $k_4 =
\partial_{[\psi]}$).  For the computation of the bulk contributions to
the index, as reported in section~\ref{ss:bulk_example}, it is not
very important to make this isometry manifest and one can use an (almost)
diagonal inverse vierbein based on the metrics \eqn{e:taub-nut_metric} or
\eqn{e:dual_metric}. It is, however, crucial
to make this symmetry manifest in
order to use the $\eta$ invariant computations for manifolds with
$S^3$ boundary of subsection~\ref{ss:eta_s3}. To achieve this one makes
use of the left-invariant forms $\Sigma^i$. These forms satisfy
\begin{equation}
{\cal L}_{k} \Sigma^{i} = \big( \iota_{k} {\rm d} + {\rm
d}\iota_{k} \big) \, \Sigma^i = \big( k^\mu \partial_\mu \Sigma^i_\nu 
+ \partial_\nu k^\mu \Sigma^i_\mu\big)\, {\rm d}x^\nu = 0\, ,
\end{equation}
where ${\cal L}_k$ is the Lie derivative with respect to any one of
the three Killing vector fields $k_{i}$. Explicitly,
\begin{equation}
\begin{aligned}
\Sigma_\mu{}^{1}\,{\rm d}x^\mu = \Sigma^1 &= \hphantom{-}\cos\psi\, {\rm d}\theta + \sin\psi\sin\theta\, {\rm
d}\phi\, ,\\
\Sigma^2 &= -\sin\psi\, {\rm d}\theta + \cos\psi\sin\theta\, {\rm
d}\phi \, ,\\
\Sigma^3 &= \hphantom{-}\cos\theta\, {\rm d}\phi + {\rm d}\psi\, ,
\end{aligned}
\end{equation}
and they satisfy ${\rm d}\Sigma^k = -(1/2)\epsilon^{ijk}
\Sigma^i\wedge \Sigma^j$. The metric \eqn{e:taub-nut_metric} of
Taub--NUT then becomes
\begin{equation}
{\rm d}\tilde s^2 = \frac{r+2m}{r}\,{\rm d}r^2 + r(r+2m)\, \left[
    (\Sigma^1)^2 + (\Sigma^2)^2\right] + \frac{4r m^2}{r+2m}\,\left[
    \Sigma^3 \right]^2\, 
\end{equation}
whose determinant is given by
\begin{equation}
\label{e:blooze}
g(\tilde e)\, = \, \Big[\,2m r (r+2m)\sin\theta\,\Big]^2
\end{equation}
The Killing--Yano tensor can also be expressed using the invariant
one-forms,
\begin{equation}
\widetilde Y \equiv \tilde e_\mu{}^r e_r{}^\rho g(\tilde e)_{\rho \nu}\,
{\rm d}x^\mu\wedge {\rm d}x^\nu= -2m\,{\rm d}r\wedge\Sigma^3 +
\frac{1}{m}(r+2m)(r+m)\,\Sigma^1\wedge\Sigma^2\, ,
\end{equation}
and is therefore $SO(3)$ invariant as well. The dual Taub--NUT
manifold, obtained through the procedure sketched in the present
section, is given by the metric
\begin{equation}
{\rm d}s^2 = 
    \frac{r+2m}{r}\, {\rm d}r^2
   +\frac{r(r+2m)}{S^2} \left[ (\Sigma^1)^2 +
    (\Sigma^2)^2 \right] 
   + \frac{4rm^2}{r+2m} \left[ \Sigma^3 \right]^2\, .
\end{equation}
with determinant
\begin{equation}
g(e)\, = \, \Big[\,\frac{2 m r (r+2m) \sin\theta}{S^2}\,\Big]^2
\end{equation}
where $S=(r+m)/m$.

\begin{table}
\begin{small}
\begin{equation*}
\begin{aligned}
\tilde e^1 = \sqrt{\frac{r+2m}{r}}\, {\rm d}r\, ,\quad &
\tilde e^2 = \sqrt{r^2+2mr} \, \Sigma^1\, ,\\
\tilde e^3 = \sqrt{r^2+2mr} \, \Sigma^2\, ,\quad &
\tilde e^4 = \sqrt{\frac{4rm^2}{r+2m}}\, \Sigma^3 \, .
\end{aligned}
\end{equation*}
\begin{equation*}
\begin{aligned}
\widetilde\omega_{12} &= -\frac{r+m}{r+2m}\, \Sigma^1\, , \quad & 
\widetilde\omega_{13} &= -\frac{r+m}{r+2m}\, \Sigma^2\, ,\\
\widetilde\omega_{14} &= -2\frac{m^2}{(r+2m)^2}\,  \Sigma^3\, , \quad &
\widetilde\omega_{23} &= -\frac{r^2+2m^2+4rm}{(r+2m)^2}\, \Sigma^3\, ,\\
\widetilde\omega_{24} &= \frac{m}{r+2m}\, \Sigma^2\, ,\quad &
\widetilde\omega_{34} &= -\frac{m}{r+2m}\, \Sigma^1\, .
\end{aligned}
\end{equation*}
\begin{equation*}
\begin{aligned}
-\widetilde R_{43} &= \widetilde R_{21} &= \frac{m}{(r+2m)^2}\, {\rm d}r\wedge \Sigma^1 
         -\frac{2rm^2}{(r+2m)^3}\, \Sigma^2\wedge \Sigma^3\, ,\\
\widetilde R_{42} &= \widetilde R_{31} &= \frac{m}{(r+2m)^2}\, {\rm d}r\wedge \Sigma^2 
         +\frac{2rm^2}{(r+2m)^3}\, \Sigma^1\wedge \Sigma^3\, ,\\
-\widetilde R_{32} &= \widetilde R_{41} &= -\frac{4m^2}{(r+2m)^3} \, {\rm d}r\wedge \Sigma^3
         +\frac{2rm}{(r+2m)^2}\,\Sigma^1\wedge\Sigma^2\, .
\end{aligned}
\end{equation*}
\end{small}
\caption{Geometrical data of the Taub--NUT manifold.}
\label{t:taub-nut_geometry}
\end{table}

\begin{table}
\begin{small}
\begin{equation*}
\begin{aligned}
e^1 = \,\sqrt{\frac{4rm^2}{r+2m}}\, \Sigma^3\, ,\quad &
e^2 = -\frac{\sqrt{m^2r(r+2m)}}{r+m}\, \Sigma^2\, ,\\
e^3 = \frac{\sqrt{m^2r(r+2m)}}{r+m}\, \Sigma^1\, ,\quad &
e^4 = -\sqrt{\frac{r+2m}{r}}\, {\rm d}r\, .
\end{aligned}
\end{equation*}
\begin{equation*}
\begin{aligned}
\omega_{12} &= -\frac{r+m}{r+2m}\, \Sigma^1\, ,\quad &
\omega_{13} &= -\frac{r+m}{r+2m}\, \Sigma^2\, ,\quad &
A_{12} &= \frac{1}{3}\frac{r}{r+m}\,  \Sigma^1\, ,\\
\omega_{14} &= -2\frac{m^2}{(r+2m)^2}\,  \Sigma^3\, , \quad &
\omega_{23} &= \frac{r^2-2m^2}{(r+2m)^2}\, \Sigma^3\, , \quad &
A_{13} &= \frac{1}{3}\frac{r}{r+m}\, \Sigma^2\, ,\\
\omega_{24} &= \frac{m^3}{(r+2m)(r+m)^2}\, \Sigma^2\, ,\quad &
\omega_{34} &= -\frac{m^3}{(r+2m)(r+m)^2}\, \Sigma^1\, ,\quad &
A_{23} &= \frac{2}{3}\frac{r}{r+2m}\, \Sigma^3\, .
\end{aligned}
\end{equation*}
\begin{equation*}
\begin{aligned}
R_{21} &= &\frac{m}{(r+2m)^2}\, {\rm d}r\wedge \Sigma^1 
  &-& 2\,\left[ \frac{(r+m)^3}{(r+2m)^3} - \frac{m^5}{(r+m)^2(r+2m)^3}
  \right] \, \Sigma^2\wedge \Sigma^3\, , \\
R_{31} &= & \frac{m}{(r+2m)^2}\, {\rm d}r\wedge \Sigma^2 
  &+& 2 \,\left[ \frac{(r+m)^3}{(r+2m)^3} - \frac{m^5}{(r+m)^2(r+2m)^3}
  \right]\, \Sigma^1\wedge\Sigma^3\, , \\
R_{41} &= & -\frac{4m^2}{(r+2m)^3} \, {\rm d}r\wedge \Sigma^3 
  &-& \frac{2m^2r}{(r+m)(r+2m)^2}\quad \Sigma^1\wedge\Sigma^2\, ,
\end{aligned}
\end{equation*}
\begin{gather*}
\begin{aligned}
R_{43} &= - &\frac{m^3(3r+5m)}{(r+2m)^2(r+m)^3}\, {\rm d}r\wedge\Sigma^1
  + \frac{2rm^2}{(r+2m)^3}\, &\Sigma^2\wedge\Sigma^3\, \\
R_{42} &=  &\frac{m^3(3r+5m)}{(r+2m)^2(r+m)^3}\, {\rm d}r\wedge\Sigma^2
  + \frac{2rm^2}{(r+2m)^3}\, &\Sigma^1\wedge\Sigma^3\, ,
\end{aligned}\\
R_{32} =  -\frac{4m(r+m)}{(r+2m)^3}\, {\rm d}r \wedge \Sigma^3 
  + \left[ \frac{2r^2+2rm-m^2}{(r+2m)^2} +
  \frac{m^6}{(r+2m)^2(r+m)^4} \right]\, \Sigma^1\wedge\Sigma^2\, .
\end{gather*}
\end{small}
\caption{Geometrical data of the dual Taub--NUT manifold.}
\label{t:dual_geometry}
\end{table}

The connections and curvatures are all summarised in the two tables
\ref{t:taub-nut_geometry} and \ref{t:dual_geometry}. Before we discuss
the Dirac operators, let us make a few remarks about the displayed
results. First of all, the curvature of the dual manifold is not
(anti)self-dual, in contrast to Taub--NUT (note that in our
conventions $\widetilde R\,{}^{rs}= - (1/2) \epsilon^{rtsu}
\widetilde R\,{}_{su}$). 
Its asymptotic geometry is also
different: whereas Taub--NUT is asymptotically a flat metric, the dual
metric has a volume element independent of $r$ at large radius; it
tends to an infinite tube.

Of course, we must also specify the coordinate ranges for the angular
coordinates $\theta$, $\phi$ and $\psi$. It is easy to see that for
both Taub--NUT and dual Taub--NUT one has $\theta\in [0,\pi)$ and
$\phi\in[0,2\pi)$.  For the coordinate $\psi$ one finds
$\psi\in[0,4\pi)$ by ensuring the absence of the NUT singularity at
$r=0$. In detail, one can write both the Taub--NUT and dual Taub--NUT
metrics on a constant $(\theta\phi)$--slice in the vicinity of
$r=0$ as ${\rm d}s^2={\rm d}x^2+\frac{1}{4}x^2 {\rm d}\psi^2$ via the coordinate
transformation $x^2=8mr$. Requiring this be the metric of a
two-dimensional plane in polar coordinates we find the quoted result
for $\psi$.

For the Dirac operators we actually need the inverse of the invariant
one-forms. We denote them by $\Sigma_i$ just as in
subsection~\ref{ss:eta_s3}. They are the vectors commuting with the
Killing vectors,
\begin{equation}
{}[ k_i , \Sigma_j ] = 0\, ,
\end{equation} 
and in components one finds
\begin{equation}
\begin{aligned}
\Sigma_{1}{}^\mu\partial_\mu = \Sigma_1 &= \cos\psi\, \partial_\theta 
  + \frac{\sin\psi}{\sin\theta}\,\partial_\phi
  - \frac{\cos\theta\sin\psi}{\sin\theta}\, \partial_\psi \,\\
\Sigma_{2} &= -\sin\psi\,\partial_\theta
  + \frac{\cos\psi}{\sin\theta}\,\partial_\phi
  - \frac{\cos\psi\cos\theta}{\sin\theta}\,\partial_\psi\, ,\\
\Sigma_{3} &= \partial_\psi\, .
\end{aligned}
\end{equation}
In the manifestly $SO(3)$ invariant basis the 
Dirac operator on dual Taub--NUT with Killing--Yano
torsion is given by
\begin{multline}
\label{e:tn_Df_sigmas}
{\slashed{D}}(e,A) = \frac{1}{rV}\,\gamma^4\,\bigg[
   r \partial_{r} + 
\frac{1}{2}\left(\frac{2r+3m}{r+2m}-\frac{2r}{(r+m)}\right)\\[.5em]
 +\left\{ S\left(\gamma^{34} \Sigma_{1} + \gamma^{42} \Sigma_{2}\right)
 + \frac{1}{\lambda} \gamma^{14} \Sigma_{3} 
 - \gamma^{1234} \frac{\lambda^2+2}{4\lambda} \right\}\bigg]\, .
\end{multline}
It is antihermitean on the dual Taub--NUT metric and anticommutes with the
operator
\begin{multline}
\label{e:tn_D_sigmas}
\slashed{D}(\tilde e,\widetilde V) = \frac{1}{rV}\,\gamma^1\,\bigg[ 
   r \partial_{r} + \frac{1}{2}\left(\frac{2r+3m}{r+2m}-\frac{2r}{(r+m)}\right)\\[.5em]
 +\left\{\gamma^{12} \Sigma_{1} + \gamma^{13} \Sigma_{2}
 + \frac{1}{\lambda} \gamma^{14} \Sigma_{3} 
 - \gamma^{1234} \frac{\lambda^2+2}{4\lambda} \right\}\bigg]\, .
\end{multline}
The operator $\slashed{D}(\tilde e,\widetilde V)$ is antihermitean on the dual Taub--NUT manifold
but has no definite hermiticity on Taub--NUT itself where it can be regarded as
the Dirac operator coupled to the trace of the torsion
$\gamma^\mu\widetilde V_\mu=-\gamma^1/[(r+m)V]$, or in other words a purely
imaginary abelian vector field. In fact the replacement
\begin{equation}
\partial_r\rightarrow\partial_r + \frac{1}{r(r+m)}
\end{equation}
renders the operators $\slashed{D}(\tilde e,\widetilde V)$ and
$\slashed{D}(e,A)$ antihermitean on Taub--NUT and in the case of the
former one obtains $\slashed{D}(e)$, the torsion-free Taub--NUT Dirac
operator.  Note that in the above $\lambda=2m/(r+2m)$, $S=(r+m)/m$ and
$V=\sqrt{(r+2m)/r}$.

\subsection{Example: Index theorems for Taub--NUT and dual Taub--NUT}
\label{ss:example_indexsummary}
\subsubsection{Bulk}
\label{ss:bulk_example}

Using the geometrical data of the two manifolds we just presented, all
terms in~\eqn{e:johnny} can be computed and the indices can be
compared. Let us start with the bulk part. For the dual
Taub--NUT manifold there are contributions from all terms
in expression~\eqn{e:bulkresult} except for the $F\wedge F$ term since
$F={\rm d}A$ vanishes exactly for this example. 
Putting the boundary at a finite
radius $r=r_b$ we obtain the (rather unenlightening) result
\begin{multline}
\label{e:bulk_dual}
\frac{1}{24.8\pi^2}\int_{\cal M} \Big[ R_{mn}\wedge R^{nm} 
- \frac{1}{2} F\wedge F - 2\, \sqrt{g}\, D_\mu {\cal K}^\mu\Big] 
= \\[1em]
\frac{1}{12}\frac{r_b^2\, ( r_b^6 + 12 m r_b^5 + 86 r_b^4 m^2 
+ 340 m^3 r_b^3 + 753 r_b^2 m^4 + 872 r_b m^5 + 408 m^6)}{(r_b+2m)^4 (r_b+m)^4}\, .
\end{multline}
In the limit $r_b\rightarrow \infty$ this yields the number $1/12$.
If one turns the torsion off, the bulk result for dual Taub--NUT is
then simply
\begin{align}
\label{e:peter}
{\frac{1}{24.8\pi^2}\int_{\cal M} R_{mn}\wedge R^{nm}}  
&= 
\,\frac{1}{12}\frac{ r_b^2\,(  9 r_b^4 + 42 r_b^3 m+ 73 r_b^2 m^2 +
 56r_b m^3  +24 m^4)}{(r_b+2m)^4 (r_b+m)^2}\, .\\
\intertext{The analogous computation for Taub--NUT is well known \cite{hawk1,eguc3},}
\label{e:bulk_taub-nut}
{\frac{1}{24.8\pi^2} \int_{\widetilde{\cal M}} R_{mn}\wedge R^{nm}} &= 
\frac{1}{12}\frac{r_b^2 \,( r_b^2 + 8r_b m + 24 m^2 )}{(r_b+2m)^4}\, .
\end{align}
For a boundary at infinity, this produces the answer $1/12$ (or
equivalently a Pontrjagin number $2$).  For Taub--NUT it is known that
this number, which seems to lead to a fractional index, in fact gets
modified by the boundary contributions~\cite{roem1,eguc4,pope1}. Because of
the general results discussed in the first part of this section, we
expect that a similar correction will arise for the dual Taub--NUT manifold as well.

\subsubsection{$\eta$ invariant}
\label{ss:eta_example}

The boundaries of Taub--NUT and the dual Taub--NUT are $S^3$ so our general
discussion for the computation of the $\eta$ invariant in
subsection~\ref{ss:eta_s3} applies. 
{}From~\eqn{e:tn_Df_sigmas} we obtain the dual Taub--NUT boundary operator
\begin{equation}
\label{e:blondie}
{\cal B}=\begin{pmatrix} P(S,Z) & 0        \\ 
                         0      & -P(-S,Z) 
         \end{pmatrix}\, ,
\end{equation}
analysed above in~\ref{ss:eta_s3} with 
\begin{equation}
\lambda = \frac{2m}{r+2m}\, , \quad S=\frac{r+m}{m}
\end{equation}
as well as
\begin{equation}
Z=\begin{cases}
\,\,0 & \text{Killing--Yano torsion},\\[.2em]
\displaystyle\frac{r}{2m} & \text{torsion-free}.
\end{cases}
\end{equation}
(the torsion-free case is in fact the value for $Z$ at which the
$\eta$ invariant can be obtained from the one of Taub-NUT by replacing
$\lambda\rightarrow S\lambda$, but we have presented the general
formula~\eqn{e:eta_general} just in case one wants to study
the interpolation between the two cases). We then find that $\eta(0)$ with the
boundary at finite radius $r=r_b$ for dual Taub--NUT with
Killing--Yano torsion is
\begin{equation}
\label{e:eta_dual}
\frac{1}{2}\eta_B(0) = -\frac{1}{12}\frac{r_b(r_b^5+10m r_b^4 + 65 r_b^3 m^2 + 168 m^3 r_b^2
+ 168 r_b m^4 + 48 m^5)}{(r_b+2m)^4 (r_b+m)^2} 
\end{equation}
and for dual Taub--NUT without torsion
\begin{equation}
\label{e:paul}
\frac{1}{2} \eta_{B\,(A=0)}(0) = -\frac{1}{12} \frac{(3 r_b + 4m)^2
     r_b^2}{(r_b+2m)^4}\, .
\end{equation}
For Taub--NUT itself, from~\eqn{e:tn_D_sigmas} the boundary Dirac operator is
\begin{equation}
\label{e:paul1}
\widetilde{\cal B}=\begin{pmatrix} P(1,0) & 0        \\ 
                         0      & -P(1,0) 
         \end{pmatrix}\, \, .
\end{equation}
Again $\lambda=2m/(r+2m)$
where this result as well as~\eqn{e:blondie} are given in the same chiral 
basis~\eqn{e:what's in a basis?} to facilitate comparison of the space of
boundary solutions with positive eigenvalues in subsection~\ref{ss:equality} below.
Clearly the analysis of subsection~\ref{ss:eta_s3} applies equally
well to the Taub--NUT boundary Dirac
operator~\eqn{e:paul1} and importantly the solutions still take the
form quoted in~\eqn{e:spectrum}. Therefore, for the Taub--NUT manifold, we
reproduce the results of~\cite{eguc3} and find
\begin{equation}
\label{e:eta_taub-nut}
\frac{1}{2}\eta_{\widetilde B}(0) = -\frac{1}{12}\frac{r_b^2(4m+r_b)^2}{(r_b+2m)^4}\, .
\end{equation}
Note that the expressions~\eqn{e:eta_taub-nut} and~\eqn{e:eta_dual} go
to $-1/12$ when the boundary is moved out to infinite radius,
$r_b\rightarrow \infty$. On the other hand, the limiting value for
$\frac{1}{2}\eta(0)$ for dual Taub--NUT without torsion is $-3/4$.

Finally we note that the number of harmonic boundary spinors $h$
vanishes for Taub--NUT and dual Taub--NUT with or without torsion. 

\subsubsection{Chern--Simons correction}
\label{ss:CS_example}

For the dual of Taub--NUT, we have computed the difference between
the boundary integral at infinity for the product metric and the same
integral for the full metric to be (the contribution $A{\rm d}A$ from the abelian axial vector
field drops out since in this example $F={\rm d}A\equiv 0$)
\begin{multline}
\label{e:cs_dual}
\frac{1}{24.8\pi^2} \int_{\partial {\cal M}} \Big[ \omega^\sharp \wedge R^\sharp -
\tfrac{1}{3}\omega^\sharp \wedge\omega^\sharp \wedge\omega^\sharp -2\,{\cal K}^\sharp - \omega \wedge R +
\tfrac{1}{3}\omega\wedge\omega\wedge\omega + 2\,{\cal K} \Big] = \\
\frac{2}{3} \frac{r_b m^3(- 4r_b^4 -23 r_b^3 m -40m^2r_b^2-18r_b m^3   + 6
m^4)}{(r_b+2m)^4(r_b+m)^4}\, .
\end{multline}
Since the dual metric is already a product metric at infinity, one
expects this correction to vanish in the large $r_b$ limit, which is
indeed the case. Turning off the torsion this reduces to
\begin{equation}
\label{e:mary}
\frac{1}{24.8\pi^2}\int_{\partial {\cal M}} \Big[ \omega^\sharp \wedge R^\sharp -
\tfrac{1}{3}\omega^\sharp\wedge\omega^\sharp\wedge\omega^\sharp 
- \omega \wedge R +
\tfrac{1}{3}\omega\wedge\omega\wedge\omega \Big] = 
-\frac{2}{3} \frac{m^4 r_b^2}{(r_b+2m)^4(r_b+m)^2}\, .
\end{equation}
Finally, for Taub--NUT itself, we confirm the result of
\cite{eguc3} which in our coordinate system is
\begin{equation}
\label{e:cs_taub-nut}
\frac{1}{24.8\pi^2}\int_{\partial \widetilde{\cal M}} \Big[ \omega^\sharp \wedge R^\sharp -
\tfrac{1}{3}\omega^\sharp\wedge\omega^\sharp\wedge\omega^\sharp 
- \omega \wedge R +
\tfrac{1}{3}\omega\wedge\omega\wedge\omega \Big] = 
-\frac{2}{3} \frac{m^2 r_b^2}{(r_b+2m)^4}\, .
\end{equation}
The above expression also vanishes in the large $r_b$ limit.

\subsubsection{Equality of indices}
\label{ss:equality}

Although it is now easy to verify the equality of the indices on the
Taub--NUT and dual Taub--NUT manifolds, it would be remiss of us to do
so without first checking the validity the assumption of
subsection~\ref{ss:hermiticity} above.  Therefore we return to the
solutions presented in equation~\eqn{e:spectrum} and check that the
eigenspaces generated by the solutions with positive eigenvalues
coincide for Taub--NUT and dual Taub--NUT. Clearly the eigenvalues
(setting $Z=0$, since $Z=r/(2m)\neq 0$ corresponds to dual Taub--NUT without
torsion) $p/2\lambda+\lambda/4$ and $\lambda/4+\sqrt{4pq(\lambda
S)^2+(p-q)^2}/2\lambda$ ($S=1$ for Taub--NUT and $S=(r+m)/m$ for dual
Taub--NUT) are always positive.  Furthermore, from the form of the
explicit solutions, we see that the eigenspaces generated by the
corresponding eigenvectors are identical for both
manifolds. (Strictly, one must take $r_b$ sufficiently small for the preceding
statement to hold. However, this is sufficient since, as we are about
to show, neither manifold possesses harmonic boundary spinors at any
positive $r_b$ so that for both manifolds, the index is independent of $r_b>0$.)

If the solutions corresponding to the negative root
$\lambda/4-\sqrt{4pq(\lambda S)^2+(p-q)^2}/2\lambda$ were all to have
strictly negative eigenvalues clearly we would be done since the
projective boundary conditions then coincide for the two
manifolds. Yet for any non-negative radius $r$ to the boundary, the
eigenvalue with the negative root can never be positive for either
manifold.  To see this we just need to look at the minimum of the
square root $\sqrt{4pq(\lambda S)^2+(p-q)^2}$ which occurs at $p=1=q$.
But for both manifolds $\lambda/4-S$ is negative so long as $r$ is
non-negative which completes our verification of the assumption.

We are now justified in comparing the indices of Taub--NUT and dual Taub--NUT.
Therefore, adding the bulk contribution~\eqn{e:bulk_dual}, the $\eta$
invariant~\eqn{e:eta_dual} and the Chern--Simons
correction~\eqn{e:cs_dual} of the dual Taub--NUT manifold, we find
that the $r_b$ dependence drops out completely,
\begin{equation}
\begin{aligned}
{\rm index}\slashed{D}(e,A)&=\quad  
\frac{r_b^8+12r_b^7m+86r_b^6m^2+340r_b^5m^3+753r_b^4m^4+872r_b^3m^5+408r_b^2m^6}
{12(r_b+2m)^4(r_b+m)^4}\\
&\quad-\,
\frac{r_b^8+12r_b^7m+86r_b^6m^2\!+308r_b^5m^3\!+569r_b^4m^4\!+552r_b^3m^5\!+264r_b^2m^6\!+48r_bm^7}
{12(r_b+2m)^4(r_b+m)^4}\\
&\quad \hphantom{-\,r_b^8+12r_b^7m+86r_b^6m^2}
-\,\frac{32r_b^5m^3+184r_b^4m^4+320r_b^3m^5+144r_b^2m^6-48r_bm^7}
{12(r_b+2m)^4(r_b+m)^4}\\
&=\,0 \, ,
\end{aligned}
\end{equation}
(the $\eta$ invariant has to be added rather than subtracted because
our conventions for the Dirac matrices imply that the upper component
states have negative chirality).  This result matches precisely the
(well known) result for Taub--NUT obtained by
adding~\eqn{e:bulk_taub-nut}, \eqn{e:eta_taub-nut} and
\eqn{e:cs_taub-nut} (it is easy to verify that there is no additional
contribution at any value of $r_b$ to the Taub--NUT index from the
abelian coupling $V^\mu$) and therefore
\begin{equation}
\label{e:miraculous_cancellations}
\text{index}\slashed{D}(e,A) = 0 =
\text{index}\slashed{D}(\tilde e,\widetilde V)\, .
\end{equation}
It is interesting to also compute the index for dual Taub--NUT without
torsion. In that case, adding~\eqn{e:peter},~\eqn{e:paul} and ~\eqn{e:mary}, we again
find a vanishing result 
\begin{equation}
{\rm index}\,\slashed{D}(e,A=0)=0
\end{equation} 
at any radius to the boundary $r_b$ (one might have even argued that 
since the dual metric can be smoothly deformed into that of Taub--NUT, this 
was to be expected, although in the presence of boundary this 
statement should be carefully re-examined).

Even though the bulk and boundary terms decouple completely in the
limit $r_b\rightarrow\infty$ for both manifolds, this is not true for
general values of $r_b$ indicating the
non-trivial nature of our result. 
Moreover, the torsion terms in the various contributions
yield a nontrivial answer which matches precisely with that of the metric
terms to produce the $r_b$ independent result
\eqn{e:miraculous_cancellations}.
This concludes our check of the APS index theorem generalised to
manifolds with torsion.

\section{Summary and conclusions}
\label{s:conclusions}

In this paper we have 
presented the modifications of the bulk and boundary terms in
the index theorems for manifolds with boundary when torsion is
added. In addition we have shown how the index theorem for such 
manifolds can be related when they possess mutually anticommuting
Dirac operators. All these results can be explicitly verified for the
case of the Taub--NUT and dual Taub--NUT manifolds and we find that
the index for both these manifolds vanishes.

Many important technical and physical issues were solved {\em en route}\/ to
the above results. In particular the appearance of the Nieh--Yan
tensor at order ${\cal O}(\beta^{-1})$ in the bulk index computation
has previously caused some 
controversy~\cite{obuk1,soo1,miel1,chan4,chan5,chan6}.
However, once one studies well-posed index problems by carefully
imposing boundary terms for spinors, the index should be independent
of the inverse regulator $(\mbox{mass})^2$ $\beta$. Therefore one
suspects that keeping all regulator dependent terms in the APS analysis
in the presence of torsion, all  $1/\beta$ terms should exactly
cancel. The importance of this remark for the axial anomaly
in quantum field theory should not go unnoticed.

Interestingly enough, we note that the torsion does not lead to a
modification of the index for the dual Taub-Nut manifold, so that
one may wonder whether on general grounds torsion should be expected
contribute to the index or not. We are not aware of any general
argument suggesting that torsion cannot contribute to the index of
manifolds {\em with boundary}\/ but from a physical viewpoint, where each of
the bulk, boundary and generalised Chern--Simons corrections are
separately of considerable interest, the existence of such an argument
would not diminish the importance of our results which we now
summarise.

The relation found between manifolds with anticommuting Dirac
operators involves the so-called Killing--Yano duality. In other
words, one must must study the motion of spinning particles in curved space
and in this paper we have extended existing studies of classical
spinning particle dynamics to the quantum case in backgrounds with torsion. 
The precise path integral quantisation was given in
section~\ref{s:bulk} and the resolution of possible ordering
ambiguities for the Dirac operator by invoking anti-hermiticity
was discussed in
section~\ref{s:dual_operators}. As mentioned earlier, our work may
also be viewed as a stringent check of the precise path integral
quantisation scheme utilised in this paper.

The relationship we found between manifolds with mutually
anticommuting Dirac operators depended on being able to understand
and compare the Hilbert space of spinors on manifolds with boundary. This
analysis was made possible by the work of APS and we have shown that
it can be equally well applied to the torsion-full case also. In
particular we note that in all our analysis it was possible to take
any {\em finite}\/ radius $r_b$ to the boundary. This allows one to
study the non-compact limit in which $r_b\rightarrow \infty$ which
of course has some topical significance in present day studies of anti de
Sitter metrics. The agreement we found between indices at finite
$r_b$ is a very strong check of our results.

The generalisations to include torsion that we presented of Hitchin's $\eta$ invariant
computation for squashed $S^3$ metrics and to Gilkey's
Chern--Simons non-product metric boundary correction were in principle
straightforward but are of course an important step if one is to
understand index theorems in the torsion-full case. We note that it
would be desirable also to consider cases where the antisymmetric
contortion also has non-vanishing components in the direction normal
to the boundary. One might expect then that analogous results to ours would
also hold. One might also like to proceed in such a case by searching
for other manifolds satisfying the Killing--Yano duality relation. To
this end we note that the Kerr--Newman metric of a rotating black hole
is certainly tractable along the lines presented in this paper, but we
reserve this physically interesting metric for further study.

As a final note we also observe that the generalised Chern--Simons form $C=\omega\wedge
R-\frac{1}{3}\omega\wedge\omega\wedge\omega-2\,{\cal K}$ found in
section~\ref{s:bulk} (where, presumably one ought include the order
$\hbar^{-1}$ Nieh--Yan term along with possible higher $O(\hbar)$
corrections in ${\cal K}$) might represent an
interesting three dimensional field theory in its own right. But again
we leave such developments to the future.

\section*{Acknowledgments}

We thank Jan-Willem van Holten for encouraging us to study this
problem and indicating to us that a relation between index theorems
for manifolds with Killing--Yano tensors and their dual manifolds
should exist.  Furthermore, we are honoured to thank Gary Gibbons,
Stephen Hawking, Sven Moch, Diana Vaman and Peter van Nieuwenhuizen
for discussions and are deeply indebted to Nigel Hitchin for providing
us the derivation of the residues $f\big((s+1)/2\big)$ and
$f\big((s+3)/2\big)$ required in the $\eta$ invariant computation~\cite{hitc2}.
The calculations reported in this paper made extensive use of the
tensor algebra package GRtensor \cite{e_grte1} along with the
algebraic manipulation program FORM~\cite{b_verm1}.

\appendix
\section{Conventions and general relativity with torsion}
\label{a:conventions}

We work exclusively with manifolds of Euclidean signature and our
Dirac matrices $\gamma^r$ are hermitean and satisfy
$\{\gamma^r,\gamma^s\}= 2\,\delta^{rs}$.  Flat (tangent space) indices
are denoted by the lower case Roman alphabet whereas curved indices
are members of the Greek alphabet.  Products of Dirac matrices are
denoted as $\gamma^{r_1\cdots r_n}=\gamma^{[r_1}\cdots\gamma^{r_n]}$
where we (anti) symmetrise with unit weight and
$\gamma^5=\frac{1}{24}\epsilon^{rstu}\gamma_{rstu}=\gamma^{1234}=\gamma^5{}^\dagger$.
We will often employ differential form notation in which ${\rm d}={\rm
d}x^\mu\partial_\mu$ and flat $SO(4)$ indices are usually suppressed
and understood to be traced over. For a useful review of gravity with
torsion see \cite{hehl1}.

Our convention for the spin connection in the presence of torsion is encapsulated in the
vierbein postulate,
\begin{equation}
\label{e:vierbeinpost}
0=\partial_\mu e_\nu{}^r - \Omega^\lambda{}_{\mu\nu} e_\lambda{}^r +  
\omega_\mu{}^r{}_s e_\nu{}^s\, ,
\end{equation}
where the contortion $K^\lambda{}_{\mu\nu}$ is given by
\begin{equation}
\Omega^\lambda{}_{\mu\nu} = \Gamma^\lambda{}_{\mu\nu} -  
K^\lambda{}_{\mu\nu}\, ,\quad\omega_\mu{}^m{}_n = \omega(e)_\mu{}^m{}_n - K^m{}_{\mu n}\, .
\end{equation}
In general, this differs from the torsion itself,
\begin{equation}
T^m = {\rm d}e^m + \omega_\mu{}^m{}_n\wedge e^n\, ,\quad
\text{or} \quad T^\rho{}_{\mu\nu} = \Omega^\rho{}_{[\mu\nu]}
\end{equation}
the precise relation being
\begin{equation}
K^\rho{}_{\mu\nu} = g^{\rho\lambda} \left(
 - T_{\lambda\mu\nu} + T_{\mu\nu\lambda} - T_{\nu\lambda\mu} \right)\,.
\end{equation}
The torsion-free spin connection is expressed in terms of the vierbein by
\begin{equation}
\label{e:spinconn}
\omega_{\mu\, rs}(e) = \tfrac{1}{2}\big[ -\lambda_{rs \mu} +
\lambda_{\mu rs} - \lambda_{\mu sr}\big]\, ,
\quad \lambda_{\mu \nu r} = 2\,\partial_{[\mu} e_{\nu] r}\, .
\end{equation}
We define the field strength of the contortion by
\begin{equation}
\label{e:contfieldstrength}
F_{\mu\nu\rho\sigma} = 4\, \partial_{[\mu} K_{\nu\rho\sigma]}\, .
\end{equation}
Our conventions for the Riemann curvature are 
\begin{align}
R(g,T)_{\mu\nu\rho\sigma} &= g_{\sigma\lambda}\, \partial_\nu \Omega^\lambda{}_{\mu\rho}
                             + \Omega_{\sigma\nu\lambda} \Omega^\lambda{}_{\mu\rho}
    - (\mu\leftrightarrow\nu) \, ,\\
\label{e:Riemannform}
R(\omega)_{\mu\nu mn} &= \partial_{\mu} \omega_{\nu mn}
 + \omega_{\mu mp}\omega_{\nu}{}^p{}_n - (\mu\leftrightarrow\nu) \, . 
\end{align}
and $R^{mn}={\rm d}\omega^{mn}+\omega^{m}{}_p\wedge\omega^{pn}=
\frac{1}{2}R_{\mu\nu}{}^{mn}{\rm d}x^\mu\wedge {\rm d}x^\nu$ (the
field strength $F$ is similarly normalised to be $F={\rm
d}A=\frac{1}{2}F_{\mu\nu}{\rm d}x^\mu\wedge {\rm d}x^\nu$). Using
\eqn{e:vierbeinpost} one verifies that
\begin{equation}
R(g,T)_{\mu\nu \rho\sigma}e_m{}^\rho e_n{}^\sigma = R(\omega)_{\mu\nu mn}\, .
\end{equation}
Our sign convention for the Ricci tensor is $R_{\mu\sigma} = R_\mu{}^\rho{}_{\rho\sigma}$.

We will mostly be dealing with torsion that is fully anti-symmetric, which
we denote by writing the symbol $A^\rho{}_{\mu\nu}$ instead of $K^\rho{}_{\mu\nu}$.
In this case, the torsion equals minus the contortion, so we will not introduce
a separate symbol for it. The axial vector obtained by dualising $A_{\rho\mu\nu}$
is denoted $A^\mu$
\begin{equation}
\label{e:torspseudo}
 A^\mu = \frac{1}{\sqrt{g}}\,\epsilon^{\mu\nu\rho\sigma} A_{\nu\rho\sigma}\, ,\quad
A_{\mu\nu\rho} = -\tfrac{1}{6}\sqrt{g}\,\epsilon_{\mu\nu\rho\lambda}  A^\lambda\, ,\quad
A^{\mu\nu\rho}A_{\mu\nu\rho} = \tfrac{1}{6}  A^\lambda  A_\lambda\, ,\quad
 F_{\mu\nu} = 2\, \partial_{[\mu}  A_{\nu]}\, ,
\end{equation}
from which one obtains 
\begin{equation}
F_{\mu\nu\rho\sigma} 
= -\tfrac{2}{3} \partial_{[\mu} (\sqrt{g}\,\epsilon_{\nu\rho\sigma]\lambda}A^\lambda)
= \tfrac{1}{6} \sqrt{g}\,\epsilon_{\mu\nu\rho\sigma}D_\lambda  A^\lambda\, ,
\end{equation}
using \eqn{e:contfieldstrength} and the Schouten identity.
For the contractions of the Riemann tensor, it is useful to separate
the pieces that only involve a Christoffel connection. When the torsion is
fully anti-symmetric, this leads to reasonably compact expressions,
\begin{align}
R(g,A)_{\mu\nu\rho\sigma} &= R(g)_{\mu\nu\rho\sigma} + 2\, D_{[\mu} A_{\nu]\rho\sigma} 
     + 2\, A^\lambda{}_{\rho[\mu} A_{\nu]\sigma\lambda} \, ,\\
R(g,A)_{\mu\sigma} &= R(g)_{\mu\sigma} 
     + D_\lambda A^\lambda{}_{\mu\sigma}
     + A^{\lambda\kappa}{}_\mu A_{\lambda\kappa\sigma}\, ,\\
R(g,A) &= R(g) 
  + A^{\lambda\kappa\delta} A_{\lambda\kappa\delta}\, .
\end{align}
$D_\mu$ denotes the covariant derivative {\em without\/} torsion. 
Expressed in terms of the torsion pseudo vector we then have (along
with analogous formulae for the Ricci and scalar curvatures)
\begin{equation}
\begin{aligned}
\label{e:RRptors}
R(g,A)_{\mu\nu\rho\sigma} &= R(g)_{\mu\nu\rho\sigma} 
 - \tfrac{1}{3} D_{[\mu}  A^\phi
  \sqrt{g}\,\epsilon_{\nu]\rho\sigma\phi}  \\[1em]
&\quad + \tfrac{1}{18} \big[ g_{\mu[\rho}  A_{\sigma]}  A_\nu
 - g_{\nu[\rho}  A_{\sigma]}  A_\mu 
 - g_{\mu[\rho} g_{\sigma]\nu}  A^\lambda  A_\lambda \big]\, .\\
\end{aligned}
\end{equation}
Finally we note that although the spin connection, in all generality
has 24 components which decompose under $SO(4)$ as a $\underline{4}$
(the trace), a $\underline{16}$ (terms with mixed symmetry) and a
$\underline{4}$ (the totally antisymmetric piece), the Dirac operator
can only couple via the Dirac matrices to the two
$\underline{4}$'s which one may interpret as the coupling to abelian
vector and axial vector fields $\widetilde V_\mu$ and $A_\mu$
respectively. Therefore the
most general Dirac operator is given by
$$\slashed{D}(e,V,A)=\gamma^r e_r{}^\mu(\partial_\mu+\frac{1}{4}\omega(e)_{\mu st}\gamma^{st}
+\widetilde V_\mu+\frac{1}{4}A_\mu\gamma^5)$$
where the spin connection with torsion is then
$$
\omega_{\mu rs}=\omega(e)_{\mu
rs}+\frac{4}{3}e_{\mu [r}V_{s]}-\frac{1}{6}\sqrt{g(e)}\, \epsilon_{\mu\nu\rho\sigma}
e^\nu{}_r e^\rho{}_s A^\sigma\, . 
$$
and $\omega(e)_{\mu rs}$ and $g(e)$ are the torsion-less spin
connection and metric determinant computed from the vierbein
$e_\mu{}^r$.  We abbreviate $\slashed{D}(e,V,A)$ to
$\slashed{D}(e,\widetilde V)=\slashed{D}(e,V,0)$,
$\slashed{D}(e,A)=\slashed{D}(e,0,A)$ or even just $\slashed{D}$, but
our intention should always be clear from the context.

\bibliographystyle{JHEP}
\bibliography{aps_torsion_revised}
\end{document}